\documentclass[preprint,12pt]{elsarticle}

\usepackage{amssymb}

\usepackage{amsmath}
\usepackage{xcolor}
\usepackage[normalem]{ulem}
\usepackage{lineno}
\DeclareMathOperator*{\argminA}{arg\,min}

\newcommand{\boldface}[1]{\boldsymbol{#1}}  %

\newcommand{\bfe}{\boldface{e}}

\newcommand{\bfk}{\boldface{k}}

\newcommand{\bfn}{\boldface{n}}

\newcommand{\bfx}{\boldface{x}}

\newcommand{\bfTheta}{\boldsymbol{\Theta}}

\newcommand{\calD}{\mathcal{D}}

\newcommand{\calF}{\mathcal{F}}

\newcommand{\calS}{\mathcal{S}}

\newcommand{\partderiv}[2]{\frac{\partial #1}{\partial #2}}

\newcommand{\Rset}{\mathbb{R}}

\newlength{\boxwidth}
\def\btheorem{\begin{theorem}}
\def\etheorem{\end{theorem}}
\def\blemma{\begin{lemma}}
\def\elemma{\end{lemma}}
\def\bproposition{\begin{proposition}}
\def\eproposition{\end{proposition}}
\def\bcorollary{\begin{corollary}}
\def\ecorollary{\end{corollary}}
\def\bdefinition{\begin{definition}}
\def\edefinition{\end{definition}}
\def\bexample{\begin{example}}
\def\eexample{\end{example}}
\def\bremark{\begin{remark}}
\def\eremark{\end{remark}}

\newcommand{\be}{\begin{equation}}
\newcommand{\ee}{\end{equation}}
\newcommand{\beq}{\begin{eqnarray}}
\newcommand{\eeq}{\end{eqnarray}}
\newcommand{\bem}{\begin{multline}}
\newcommand{\eem}{\end{multline}}
\newcommand{\ba}{\begin{align}}
\newcommand{\ea}{\end{align}}

\renewcommand{\figurename}{Fig.}

\journal{Extreme Mechanics Letters}

\begin{document}

\begin{frontmatter}

\title{Experiment-Informed Finite-Strain Inverse Design
of Spinodal Metamaterials}

\author[1]{Prakash Thakolkaran \fnref{+}} %
\author[2]{Michael Espinal\fnref{+}}
\author[2]{Somayajulu Dhulipala}
\author[1]{Siddhant Kumar \corref{*}\fnref{-}}

\fntext[+]{These authors contributed equally to this work.}
\fntext[-]{These authors contributed equally to this work.}
\cortext[*]{Corresponding authors}
\ead{sid.kumar@tudelft.nl}
\author[2]{Carlos M. Portela \corref{*}\fnref{-}}
\ead{cportela@mit.edu}
\affiliation[1]{organization={Department of Materials Science and Engineering, Delft University of Technology},%
            addressline={2628 CD Delft}, 
            country={The Netherlands}}
            
\affiliation[2]{organization={Department of Mechanical Engineering, Massachusetts Institute of Technology},
            addressline={77 Massachusetts Avenue},
            city={Cambridge},
            state={MA 02139},
            country={USA}}

\begin{abstract}
Spinodal metamaterials, with architectures inspired by natural phase-separ\-ation processes, have presented a significant alternative to periodic and symmetric morphologies when designing mechanical metamaterials with extreme performance. While their elastic mechanical properties have been systematically determined, their large-deformation, nonlinear responses have been challenging to predict and design, in part due to limited data sets and the need for complex nonlinear simulations. This work presents a novel physics-enhanced machine learning (ML) and optimization framework tailored to address the challenges of designing intricate spinodal metamaterials with customized mechanical properties in large-deformation scenarios where computational modeling is restrictive and experimental data is sparse. By utilizing large-deformation experimental data directly, this approach facilitates the inverse design of spinodal structures with precise finite-strain mechanical responses. The framework sheds light on instability-induced pattern formation in spinodal metamaterials---observed experimentally and in selected nonlinear simulations---leveraging physics-based inductive biases in the form of nonconvex energetic potentials. Altogether, this combined ML, experimental, and computational effort provides a route for efficient and accurate design of complex spinodal metamaterials for large-deformation scenarios where energy absorption and prediction of nonlinear failure mechanisms is essential.
\end{abstract}


\end{frontmatter}

\section{Introduction}
\label{Intro}

The rapid advancement of resolution and throughput in additive manufacturing has opened doors to new possibilities in engineering mechanical metamaterials (or architected materials) with properties that were once unattainable using conventional manufacturing techniques \cite{buckmann2012tailored,meza2014strong,meza2015resilient,vyatskikh2018additive,vangelatos2019architected}. These properties include, e.g., high strength-to-weight ratios \cite{zheng2014ultralight,meza2014strong}, negative Poisson's ratios \cite{zhang2022novel}, mechanical cloaking abilities \cite{wang2022mechanical}, tailorable anisotropic stiffnesses \cite{kumar2020inverse, zheng2023unifying, van2023inverse, bastek2022inverting}, and high energy absorption \cite{bauer2021tensegrity, wu2022twin}. While explorations on mechanical metamaterials have primarily focused on periodic and symmetric truss- \cite{saccone2022additive} and plate-based \cite{tancogne20183d} lattices, recently more attention has been given to shell-based morphologies. These morphologies,  such as triply periodic minimal surfaces  \cite{al2018microarchitected,bonatti2019smooth} and spinodal architectures \cite{kumar2020inverse, portela2020extreme,vidyasagar2018microstructural,hsieh2019mechanical} have gained traction in the field as they do not possess nodes or joints, behave in a mechanically efficient manner due to their doubly curved shells, and inherently mitigate stress concentrations. 

\begin{figure}[t!]
	\includegraphics[width = 1.0\textwidth]{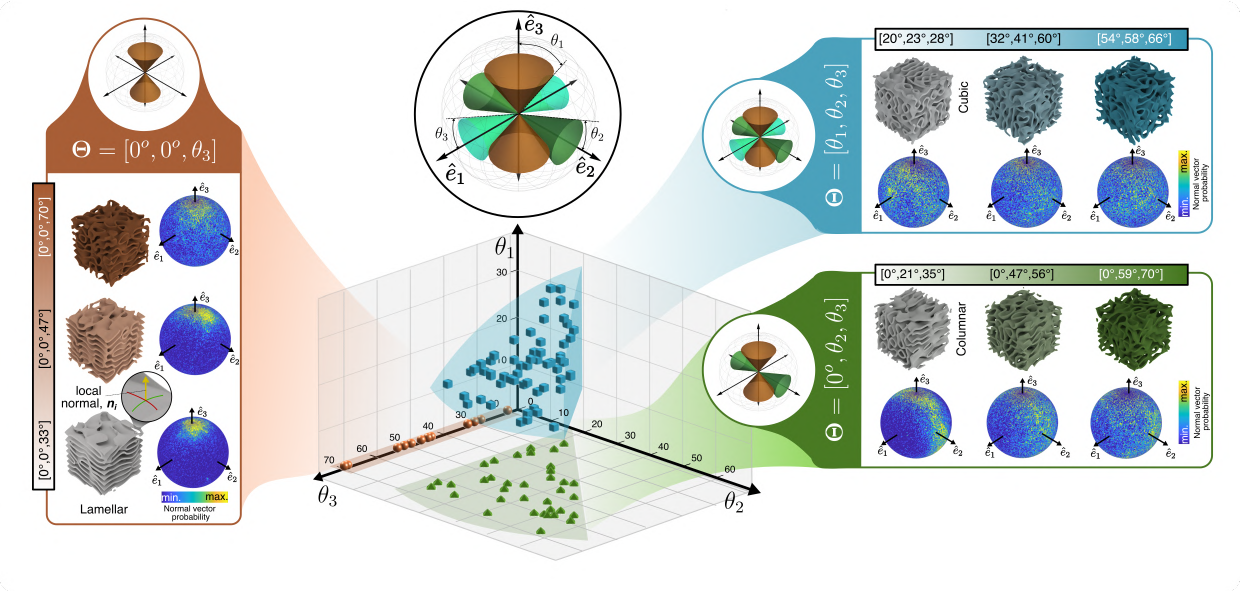}
	\centering
	\caption{Spinodal morphology design space defined by a three-parameter $\bfTheta$ representation (center). Three design subspaces, defined by the non-zero dimensionality of the $\bfTheta = [\theta_1,\theta_2,\theta_3]$ vector are represented by color clouds encompassing the designs used for training. The one-dimensional (1D) subspace corresponding to \emph{lamellar} morphologies was represented by a non-zero $\theta_3$, while the 2D and 3D subspaces subsequently added non-zero $\theta_2$ and $\theta_3$ parameters and corresponded to \emph{columnar} and \emph{cubic} morphologies, respectively. Three representative designs of increasing norm $|\bfTheta|$ are presented within each subspace, along with pole figures denoting the directional probability of normal vectors $\bfn$.} %
	\label{fig:designSpace}
\end{figure}

Spinodal-like---or \textit{spinodoid}---metamaterials are especially intriguing because they possess an aperiodic and asymmetric microstructure resembling the morphologies observed during the early stages of spinodal decomposition or rapid diffusion-driven phase separation in a homogeneous mixture of immiscible phases (\figurename~\ref{fig:designSpace}). These nature-inspired designs spanning multiple length scales---from nanoscale to macroscale---can either be manufactured using scalable self-assembly via polymerization-induced phase separation in polymer blends \cite{portela2020extreme} or additive manufacturing of morphologies extracted from phase separation simulations \cite{vidyasagar2018microstructural,hsieh2019mechanical}. In contrast to truss-, plate- and shell-based lattices, the resulting smooth and bicontinuous topologies enable robustness to manufacturing defects, mitigate any harmful stress concentrations, and exhibit extreme mechanical resilience \cite{portela2020extreme,hsieh2019mechanical}. \emph{In silico} tuning of the underlying energetics of the spinodal decomposition process opens up a diverse design space of topologies and corresponding mechanical properties \cite{vidyasagar2018microstructural,kumar2020inverse,guo2023inverse}, leading to a recent flurry of proposed spinodal metamaterials for ultralight structures \cite{bauer2022nanoarchitected,zheng2021data,portela2020extreme,senhoraSpinodal2022}, energy absorption \cite{zhang2021mechanical,guell2019ultrahigh,hsieh2019mechanical}, bone-mimetic implants \cite{kumar2020inverse,hsieh2021architected,Deng2024AIEnabled}, acoustics \cite{wojciechowski2023additively}, mass transport \cite{roding2022inverse}, among other applications.

Despite the recent advances, the structure-property relations of spinodal metamaterials have primarily been explored only in a forward fashion---computational or experimental mechanical analyses of a few representative samples chosen through trial-and-error or intuition of the design space \cite{bauer2022nanoarchitected,zheng2021data,portela2020extreme,zhang2021mechanical,guell2019ultrahigh,hsieh2019mechanical,soyarslan20183d}.
Furthermore, clear links between specific features of the morphology and the ensuing mechanical response are lacking. In light of the vast design space of spinodal metamaterials, the question of interest is that of inverse design, i.e., how to efficiently identify designs with tailored properties dictated by stress-strain responses.
While machine learning (ML) has recently seen some success in inverse design of metamaterials \cite{kumar2020inverse,zheng2021data,bastek2022inverting,zheng2023unifying,guo2023inverse,bastek2023inverse,ha2023rapid,van2023inverse, lee2023deep,lee2023data,Wilt2020,Mao2020,Meyer2022}---including but not limited to spinodal metamaterials---these algorithms face a challenge of quality-quantity duality.

High-throughput simulations of mechanical responses yield large quantities of data, which are of sufficient fidelity only in linear and small-strain regimes. For instance, in the context of spinodal metamaterials, ML models have been trained on \emph{in silico} data to inversely design for linear stiffness tensor components and tailored anisotropy \cite{kumar2020inverse}. However, in the case of finite-strain behaviors, responses due to phenomena such as material and geometric nonlinearities, buckling, self-contact, dissipation, fracture, and meshing artifacts add significant complexity and computational cost. 
On the contrary, while experimental data can serve as the ground-truth, highest-fidelity data for ML models, the quantity of data is severely limited by the number of time-consuming and costly experiments that can be reasonably performed.

Here, we propose a direct experiment-to-ML inverse design framework to design spinodal metamaterials with tailored finite-strain mechanical responses.
We create a reasonably sized dataset of 107 shell-based spinodal metamaterial architectures along with their stress-strain responses to 40\% strain along three principal directions---efficiently obtained experimentally via \emph{ex situ} and \emph{in situ} uniaxial compression at the microscale. The ML framework consists of a forward model that surrogates the structure-property relations and an inverse optimization scheme that finds the designs for a target large-deformation stress-strain response. To address the quality-quantity duality challenge, we introduce a new ML architecture that uses physics-enhanced inductive biases to eliminate the need for large-quantity and low-quality simulation data. 
To complement these predictions, we employ nonlinear finite element models that capture architecture-dependent deformation mechanisms, and shed light on the sources of energy absorption upon large deformation. In agreement with experiments and predictions, these simulations explain the relation between the directional curvature distribution and predicted energy absorption metrics.   
Furthermore, we demonstrate the ML framework's ability to identify a spinodal morphology that exhibits a target behavior that lies outside the training data domain. Altogether, this experiment-informed ML effort closes the gap between the structure-property relations of spinodal metamaterials in a large-deformation regime, particularly by accounting for nonlinear responses due to deformation mechanisms in complex architectures.

\section{Results and Discussion}
\subsection{Spinodal morphology design space}
To replicate a variety of morphologies obtained through spinodal decomposition, such as those obtained in diffusion-driven phase separation of a homogeneous mixture of immiscible phases, our first step is to define a design space that parametrizes all possible morphologies. Cahn \cite{cahn1965phase} demonstrated that the resulting phase field solution to the canonical Cahn-Hilliard equation can be approximated by a Gaussian random field (GRF), i.e., a superposition of a large number of standing waves with a narrow band of similar wavenumbers. In Fourier space, this corresponds to a  spectral density function (SDF) given by a diffused spherical surface of radius equal to the wavenumber (denoted by $\beta$ henceforth) and centered at the origin. Inspired from the formalization of this approximation \cite{roding2022inverse,Iyer2020}, we construct the phase field $\varphi:\Omega\rightarrow\Rset$ in a domain $\Omega\subset\Rset^3$ directly in Fourier space as

\begin{equation}
\varphi = \calF^{-1}\bigg[\calF[\varphi_\text{noise}]\odot\calF[\varphi_\text{filter}]\bigg] = \calF^{-1}\bigg[\calF[\varphi_\text{noise}]\odot\left({\rho[\varphi_\text{filter}]}\right)^{1/2}\bigg].
\label{eq:fourier}
\end{equation}

Here, $\calF[\cdot]$, $\calF[\cdot]^{-1}$, and $\rho[\cdot]$ denote the Fourier transform, inverse Fourier transform, and SDF (squared magnitude of the Fourier transform), respectively, while $\odot$ denotes the Hadamard product, $\varphi_\text{noise}$ represents the initial phase field of a homogeneous mixture of immiscible phases as an independent and identically distributed standard Gaussian noise (i.e., zero mean and unit variance). Convolution of $\varphi_\text{noise}$ with $\varphi_\text{filter}$ (equivalently, Hadamard product in the Fourier space) represents the phase separation process that transforms the homogeneous mixture $\varphi_\text{noise}$ into the phase-separated phase field $\varphi$. We define $\varphi_\text{filter}$ through its SDF as 
\be
    \begin{aligned}
    \rho[\varphi_\text{filter}](\bfk) &=
    \underbrace{\exp\left(-\frac{(r-\beta)^2}{2\lambda_r^2}\right)}_{\text{wavenumber control}} 
     \odot 
    \underbrace{ \sum_{i=1}^3  \sigma\left(-\lambda_\phi (\phi_i - \theta_i)\right)}_{\text{anisotropy control}}
    \\ & \text{with} \qquad  \phi_i = \min\{\cos^{-1}(k_i/r),-\cos^{-1}(k_i/r)\} \quad \text{and} \quad r=\|\bfk\|,
    \end{aligned}
    \label{eq:sdf}
\ee
where $\bfk$ denotes a wave-vector in the Fourier space; with  $\lambda_r>0$ and $\lambda_\phi>0$ as constants. The first term---labeled \textit{wavenumber control}---specifies that the spectral density is limited to a narrow Gaussian band with mean wavenumber $\beta$ and standard deviation $\lambda_r$. Choosing a higher $\beta$ yields morphologies with a finer microstructural lengthscale. The second term---labeled \textit{anisotropy control}---adds directional constraints to the wave-vectors. Specifically, the wave-vectors' orientations (given by $\phi_i$) are constrained to cones centered at the origin and along the principal axes $\{\hat\bfe_1,\hat\bfe_2,\hat\bfe_3\}$ with half-angles $\{\theta_1,\theta_2,\theta_3\}$, respectively (\figurename~\ref{fig:designSpace}, center). However, instead of a hard limit, we relax the constraint by using a smooth sigmoid-type function, which in this case is $\sigma(\cdot)=(1+\tanh(\cdot))/2$. Consequently, the  probability of a wave-vector outside the cones decreases fast but smoothly with the rate determined by $\lambda_\phi$. We add both radial and angular smoothing in the SDF to mitigate formation of non-smooth artifacts in the generated morphology (see Supplementary Fig. S1). We highlight that this model of anisotropy control is not just a mathematical construct; rather it serves as an approximation to the canonical Cahn-Hilliard equation with anisotropic mobility \cite{kumar2020inverse}.

Choosing a cubic domain $\Omega$ of size $\ell\times\ell\times\ell$, the Fourier transforms in Eq.~\eqref{eq:fourier} are performed on a uniform grid of $\Omega$ with $X\times X\times X$ resolution, automatically ensuring triple periodicity in the generated structure. The morphology of the spinodal metamaterial is obtained by computing the zero level-set of $\varphi$, i.e., for $\bfx\in\Omega$,
followed by extruding the surface pointwise along both inward and outward normal directions equally for a final surface thickness of $h\ll\ell$. Note that due to the randomness of $\varphi_\text{noise}$, the resulting structures are stochastic and hence two morphology realizations for the same design parameters may be different. To reduce the effects of stochasticity, 
we choose $\beta\ell$ and 
$h/\ell$ to be sufficiently high and low, respectively, to ensure separation of scales between the microstructural length scale and the domain dimensions. We validate the choice in $\beta$ and $h$ through a systematic computational homogenization analysis as well as manufacturing considerations, respectively (Supplementary Note 2).

For the scope of this work, we uniquely defined each spinodal metamaterial design by the design parameters $\bfTheta = [\theta_1,\theta_2,\theta_3]$ with $\theta_1,\theta_2,\theta_3\in \{0\}\cup[\theta_\text{min},\theta_\text{max}]$ (i.e., the cone angles), while keeping the remaining parameters constant (see Supplementary Table S1). The angles, when non-zero, are lower-bounded by $\theta_\text{min}=20^\circ$ to ensure bi-continuity of the structures \cite{kumar2020inverse} and upper-bounded by $\theta_\text{max}=70^\circ$ to avoid degenerate (almost) isotropic structures.  For instance, when $\bfTheta=[0,0,\theta_3]$, this one-dimensional (1D) parameter subspace results in structures with distinctive lamellar-like features (\figurename~\ref{fig:designSpace}, left). In contrast,  the corresponding 2D and 3D parameter subspaces---$\bfTheta=[0,\theta_2,\theta_3]$ and $\bfTheta=[\theta_1,\theta_2,\theta_3]$---exhibit columnar- or cubic-like features, respectively (\figurename~\ref{fig:designSpace}, right). The resulting morphologies can be characterized by their corresponding surface-normal distributions, represented as spherical pole figures in \figurename~\ref{fig:designSpace}, indicating the directional distribution of material curvature within the morphologies. Overall, the design space admits a large and diverse set of morphological anisotropy, with the aim of linking said structures to their unique mechanical responses. 

Notably, increases in the magnitudes of the $\theta_i$ parameters, approaching the theoretical maximum of $90^\circ$,  results in isotropically distributed wave-vectors and the anisotropic structural distinctions disappear. However, for the sake of clarity in the subsequent discussions, we use the terminology of \textit{lamellar}, \textit{columnar}, and \textit{cubic} structures when referring to the 1D, 2D, and 3D subspaces of $\bfTheta$, respectively, regardless of their absolute values.

\subsection{Dataset generation via nanomechanical experiments}
\begin{figure}[t!]
	\includegraphics[width = 1.0\textwidth]{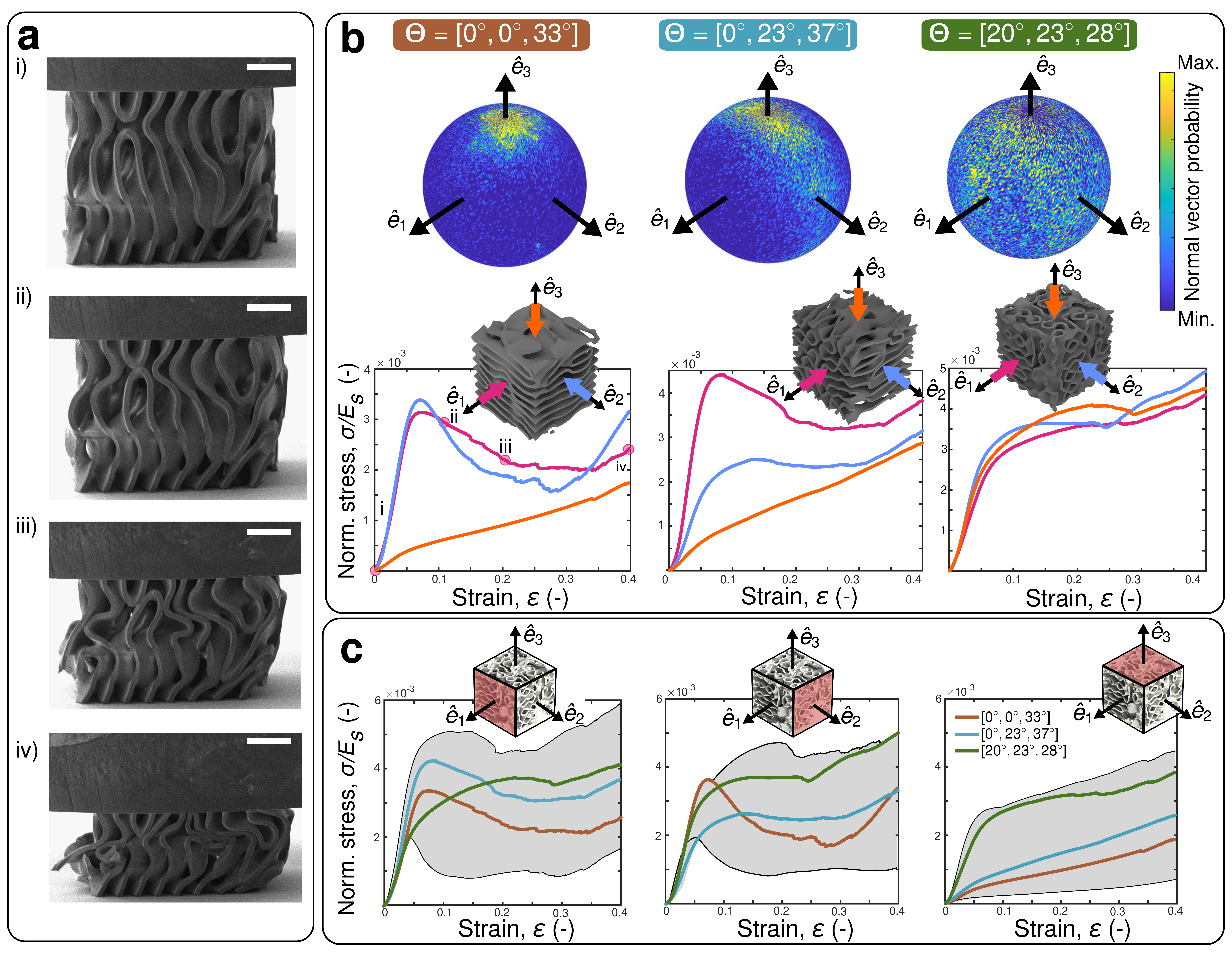}
	\centering
	\caption{Results and analysis of nanomechanical experiments. (\textbf{a}) \emph{In situ} snapshots $i$--$iv$ showing the progression of deformation for the $\bfTheta=[0^\circ,0^\circ,33^\circ]$ morphology along the $\hat{\bfe}_1$ direction, up to 40\% strain. Scale bars, 20 \textmu{}m. (\textbf{b}) Qualitative structure-to-response relations enabled by spherical pole figures denoting the directional surface-normal distributions for representative \emph{lamellar} (left), \emph{columnar} (center), and \emph{cubic} (right) morphologies---accompanied by corresponding stress-strain responses along the $\hat{\bfe}_i$ directions. The pole figures serve as a proxy for structural anisotropy, with higher surface-normal distributions along a given direction correlating to a more compliant response. The lamellar and columnar morphologies exhibited negative-stiffness regions corresponding to nonlinear buckling (as marked for the lamellar sample shown in (a)), along with stiffening at large deformations due to self-contact of shells. (\textbf{c}) Range of finite stress-strain behaviors across the training data as observed from \emph{ex situ} compressions along the three principal directions $\hat\bfe_1$ (left), $\hat\bfe_2$ (center), $\hat\bfe_3$ (right) highlighted on a generic spinodal morphology. The black lines denote the stress bounds across the training dataset, while color-coded responses correspond to the three representative morphologies shown in (b).}    
	\label{fig:Mechanics}
\end{figure}

Sampling from the design space defined above, we generated a dataset consisting of $N=107$ spinodal morphologies by randomly sampling $\bfTheta$. The dataset included 11 lamellar ($\bfTheta=[0,0,\theta_3]$), 36 columnar ($\bfTheta=[0,\theta_2,\theta_3]$), and 60 cubic ($\bfTheta=[\theta_1,\theta_2,\theta_3]$, with $\theta_1<\theta_2<\theta_3$) morphologies. To augment our experimental effort, we removed redundancy in data generation due to permutation symmetry in the parameterization (i.e., the response along $\hat{\bfe}_1$ for $[\theta_1,\theta_2,\theta_3]$ is equivalent to the response along $\hat{\bfe}_2$ for $[\theta_2,\theta_1,\theta_3]$). {To assess the response along the three principal directions for each of the unique $N$ morphologies, we fabricated 321 samples corresponding to 107 geometries, printed in three orientations. Altogether, these samples represented 609 effective $\mathbf{\Theta}$ parameterizations within our design space, accounting for permutation symmetry (see Supplementary Note 1). The 321 samples were fabricated out of IP-Dip photoresist using a two-photon lithography process (Supplementary Note 2), resulting in cubic unit cells with an average edge length of $\ell=92$ \textmu{}m and shell thickness of $h=2.6$ \textmu{}m, with an approximate relative density (i.e., fill fraction) of 40\%.} Using sub-micron resolution X-ray computed tomography, we validated the geometric validity of our fabricated samples, ensuring accurate representation of curvatures, shell thicknesses, and relative densities (Supplementary Note 2). 

{We note that, imperfections due to additive manufacturing can affect the mechanical properties~\cite{glaesener2023predicting,derveni2022postbuckling,gross2019correlation}. However, we did not aim to train the ML surrogate model on data whose imperfections have been artificially reduced through repeated experiments. Instead, we let the model account for the role of as-fabricated imperfections in the mechanical response.}

To obtain the finite-strain response of each sample in the dataset, we performed both \emph{ex situ} and \emph{in situ} quasi-static uniaxial compression experiments (strain rate of $10^{-3}$ s$^{-1}$) along the three principal directions $\hat{\bfe}_1$, $\hat{\bfe}_2$, and $\hat{\bfe}_3$. \emph{In situ} observation of the compression enabled visualization of multiple nonlinear and irreversible mechanisms such as plastic buckling, self-contact, and fracture through the thickness of the shells
(\figurename~\ref{fig:Mechanics}a)---which could be linked to specific characteristics of the large-deformation stress-strain response.  
The loading portion of the measured 321 stress-strain curves was used to train the ML model for inverse design, while the unloading portion was omitted since it carried minimal information in this large-strain regime (full loading-unloading responses are presented in Supplementary Note 1. To facilitate comparison to spinodal metamaterials made of similar polymeric constituents, we report the stress normalized by the elastic modulus $E_s$ ($3.2 \pm 0.3$ GPa) of IP-Dip photoresist, determined from experiments on monolithic micropillars with the same print parameters as the metamaterials. 

As an appropriate qualitative indicator of mechanical anisotropy within each design, we represented the surface normal distributions in the form of a spherical pole figure, indicating the relative orientation of curved shells with respect to a loading direction (\figurename~\ref{fig:Mechanics}b). Regions with a higher distribution of surface normals correlated to a more compliant response in the linear regime, followed by an on-average lower stress level throughout subsequent deformation. In the cases of lamellar and columnar morphologies, directions with low surface-normal distributions tended to exhibit plastic buckling beyond the onset of nonlinearity (at strains of $\varepsilon\approx5\%$) and consequently, negative-stiffness responses up to strains of $\varepsilon\approx20\%$ as observed via \emph{in situ} experiments on the $\bfTheta=[0^\circ,0^\circ,33^\circ]$ morphology (\figurename~\ref{fig:Mechanics}a,b left). Notably, the cubic %
morphologies approach similar responses when loading in all three directions with $\theta_1 \approx \theta_2 \approx \theta_3$.  Further \emph{in situ} observations for representative samples presented in \figurename~\ref{fig:Mechanics}b are included in Supplementary Note 2 (see Supporting Information Movies S1--S9). 

When analyzing the dataset as a whole---simply represented in  \figurename~\ref{fig:Mechanics}c as bounded by maximum and minimum responses with some highlighted morphologies within---we identify the buckling behavior to be prevalent in other lamellar morphologies as well as some low-angle columnar morphologies (i.e., when $\theta_2<30^\circ$). 
For the subset of fabricated samples, we identified a distinctively different response in $\hat{\bfe}_3$-loaded responses, primarily exhibiting monotonically increasing stress levels---a consequence of our $\theta_1<\theta_2<\theta_3$ criterion when selecting morphologies. 
We highlight that between the strains of $\varepsilon\approx20\%$ to 40\%, micro-cracks began to form causing fracture events that manifested as fluctuations in the stress response. Defining the energy absorbed as the integral of the stress-strain responses to 40\% strain provided a broad distribution of performance metrics as a function of morphology and orientation. Altogether, this comprehensive dataset sheds light on the complex nonlinear responses of high-relative-density spinodal morphologies, identifying a correlation between mechanisms such as buckling and self contact to qualitative morphology classifications. These observations add intuition to previously observed nonlinear responses in spinodal morphologies, while alternate mechanics-driven computational tools are required to identify structure-property relations in this highly nonlinear regime.

\subsection{Forward Modeling via Physics-Enhanced Deep Learning}\label{sec:ml-framework}
Learning the highly nonlinear map from the design parameters $\bfTheta$ to the direction-dependent stress-strain responses of spinodal metamaterials would require a significant amount of data. To circumvent this issue and to work with limited experimental data available, we introduce a physics-enhanced ML framework that serves as a surrogate to the forward structure-to-property relations.

In the exemplar stress-strain behaviors in \figurename \ \ref{fig:Mechanics}b, we observe nonlinear features that are typical of instabilities and pattern formation in bulk materials due to nonconvex energetics \cite{kumar2020assessment}. Therefore, we model a representative stress-strain response as the derivative of an underlying strain energy density potential, which in the context of uniaxial compression corresponds to the area under the stress-strain curve. Consequently, the potential as a function of the applied strain must be monotonically increasing while admitting nonconvexities to allow for instabilities. We emphasize that this notion of energy potential is used only as an inductive bias to facilitate the learning of the uniaxial stress-strain response; its physical and thermodynamical admissibility should not be considered as strictly as a homogenized constitutive model. 

We model the above potential with a deep neural network (NN) $W(\varepsilon,\bfTheta)$ as a function of the applied strain $\varepsilon$ and design parameters $\bfTheta$. A  classical NN based on, e.g., a multi-layer perceptron (MLP) architecture may not satisfy the constraint of monotonic increase with $\varepsilon$. Additionally, while such an NN is highly non-convex in its inputs by default,  the degree of nonconvexity (loosely speaking) with respect to $\varepsilon$ should be constrained. The experimental data clearly exhibits a single macrostructural instability, which should be accordingly reflected in the NN output. Since we are training on experimental data directly,  the NN---left unchecked---can exhibit fine-scale but highly oscillatory behavior trying to fit the experimental noise \cite{as2022mechanics}. To satisfy the above constraints, we introduce the following NN architecture.

\begin{figure}[t!]
	\includegraphics[width = 1.0\textwidth]{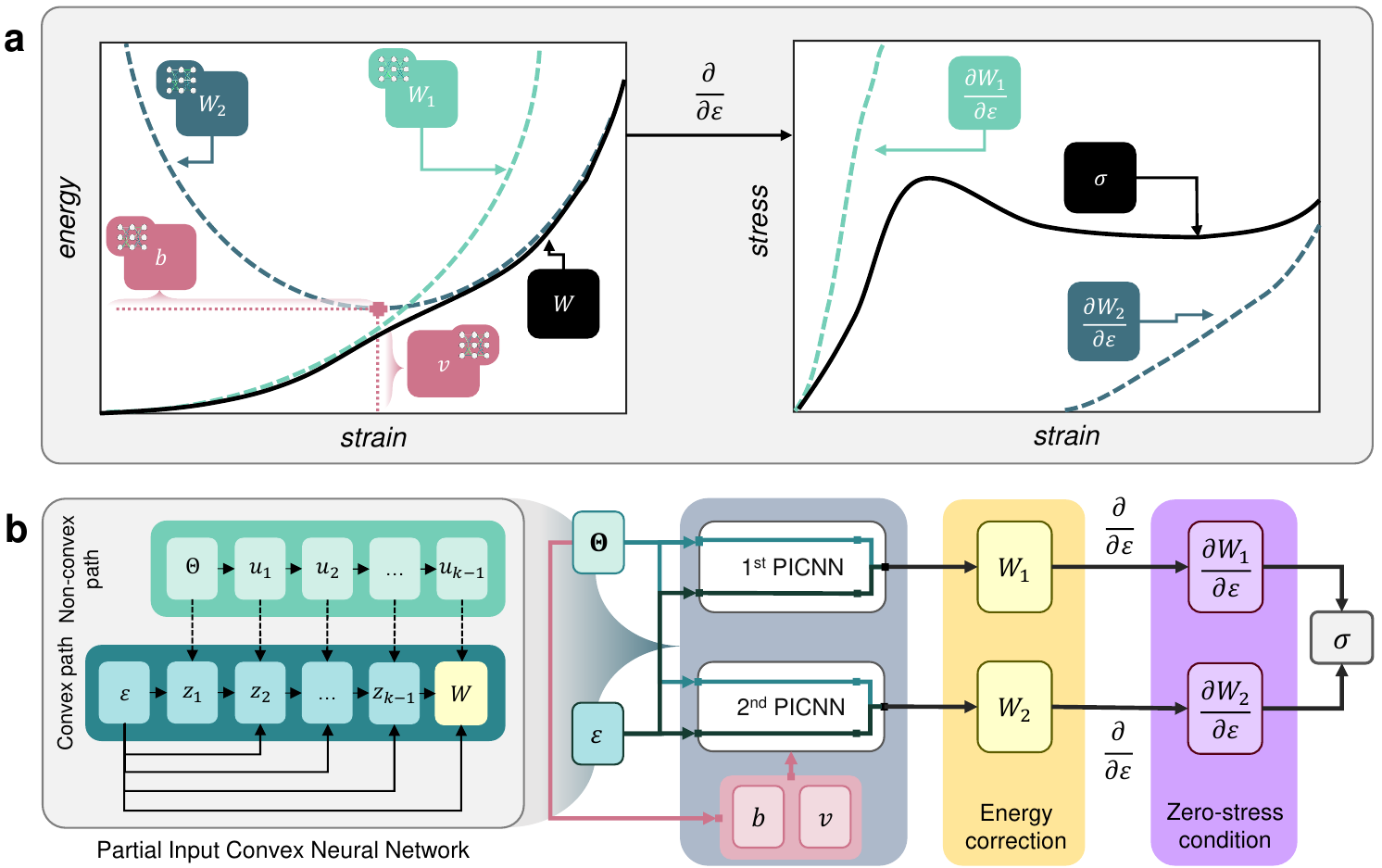}
	\centering
	\caption{Physics-enhanced deep learning framework. (\textbf{a}) \textit{Left:} The uniaxial compression stress response (as a function of applied strain $\varepsilon$) of spinodal metamaterials is modeled as the derivative of a deep neural network-based nonconvex energy density potential $W(\varepsilon,\bfTheta)$. The model consists of two potentials convex in $\varepsilon$ and given by separate neural networks: $W_1(\varepsilon,\bfTheta)$ with the energy and stress vanishing at $\varepsilon=0$ by construction; $W_2(\varepsilon,\bfTheta)$ with the energy $(v+W_1(b,\bfTheta))$ and vanishing stress at $\varepsilon=b$. Both $b(\bfTheta)$ and $v(\bfTheta)$ are also given by neural networks. The nonconvex potential $W$ is obtained by a combination of $W_1$ and $W_2$. \textit{Right:} The stress is obtained by differentiating $W$ with respect to $\varepsilon$. Also shown are the derivatives of $W_1$ and $W_2$ for reference. (\textbf{b})  Schematic of the partial input convex neural network (PICNN) architecture for $W_1$ and $W_2$ and their combination thereof. The PICNN architecture predicts an energy which is convex with respect to the strain $\varepsilon$ (via convex path) and parameterized by the design parameters $\bfTheta$ (via nonconvex path). See SI Appendix Machine Learning Framework section for details.}  %
	\label{fig:FWD_Framework}
\end{figure}

We take inspiration from Kumar et al. \cite{kumar2020assessment} and combine multiple convex potentials into one monotonically increasing, but nonconvex potential as
\be\label{eq:Wmin}
W(\varepsilon,\bfTheta) = \min\{W_1(\varepsilon,\bfTheta), W_2(\varepsilon,\bfTheta)\},
\ee
where $W_1(\varepsilon,\bfTheta)$ and $W_2(\varepsilon,\bfTheta)$ are two energy potentials that are convex in $\varepsilon$ (but non-convex in $\bfTheta$). However, the corresponding transition between phases is very sharp and not representative of the experimental data. Therefore, we relax Eq.~\eqref{eq:Wmin} by allowing the phases to coexist with volume fractions $\gamma_1,\gamma_2\in[0,1]$, respectively, as
\be
\begin{aligned}
\label{eq:Wrelax}
W(\varepsilon,\bfTheta) = & \min_{\gamma_1,\gamma_2}\  \Bigg[\sum_{i=1}^2 \gamma_i W_i(\varepsilon,\bfTheta)  - \underbrace{\left( - k_T \sum_{i=1}^2\gamma_i \log \gamma_i \right)}_{\text{configurational entropy}}\Bigg], \\ &\text{with} \quad \gamma_1 + \gamma_2 = 1.
\end{aligned}
\ee
Here, the configurational entropy (not physical entropy) penalizes the formation of phase mixtures. The constant $k_T>0$ controls the influence of configurational entropy  and in turn, the smoothness of the transition between $W_1$ and $W_2$. The effective stress $\sigma$ is obtained as the derivative of $W(\varepsilon,\bfTheta)$ 
\be
\begin{aligned}
\label{eq:sigma}
\sigma(\varepsilon,\bfTheta) = \partderiv{W}{\varepsilon} = &  \frac{\exp\left(-\frac{W_1}{k_T}\right)}{\exp\left(-\frac{W_1}{k_T}\right)+\exp\left(-\frac{W_2}{k_T}\right)} \frac{\partial W_1}{\partial\varepsilon}+\\
&\frac{\exp\left(-\frac{W_2}{k_T}\right)}{\exp\left(-\frac{W_1}{k_T}\right)+\exp\left(-\frac{W_2}{k_T}\right)} \frac{\partial W_2}{\partial\varepsilon},
\end{aligned}
\ee
where the simplification on the right side follows from the analytical solution of Eq.~\eqref{eq:Wrelax} (see Supplementary Note 3 for derivation). 

We model the constituent potential $W_1$ as 
\be
W_1(\varepsilon,\bfTheta) = 
    \underbrace{\mathcal{P}_{\omega_1}(\varepsilon,\bfTheta)}_\text{first convex potential} 
    - \underbrace{\mathcal{P}_{\omega_1}(0,\bfTheta)}_\text{energy correction}
    -\underbrace{\varepsilon\frac{\partial \mathcal{P}_{\omega_1}(0,\bfTheta)}{\partial \varepsilon}}_\text{stress correction}.
\ee
Here, $\mathcal{P}_{\omega_1}$ denotes a \textit{partial input convex NN} (PICNN) (see ref. \cite{amos2017input} for architectural details)  with the trainable parameter set $\omega_1$. Due to the inherent property of PICNNs, the predicted $W_1$ is modeled to be only convex with respect to ${\varepsilon}$, but can have any arbitrary functional relationships with $\bfTheta$. We name these functional relationships within the PICNN architecture as the convex and nonconvex path (as seen in \figurename \ \ref{fig:FWD_Framework}b). Following the principle that non-negative weighted sums of convex functions are convex, the convex path only contains linear transformations with non-negative weights and convex non-decreasing nonlinear activation functions. Supplementary Note 3 provides further details on the PICNN architecture used here. This PICNN-based approach allows us to obtain energies convex in $\varepsilon$ but non-convexly parameterized by $\bfTheta$. {Note that, $\mathcal{P}_{\omega_1}$ is merely the convex output of the PICNN, and only when combined with the energy and stress corrections it becomes the potential $W_1$.} These correction terms ensure that $W_1$ identically satisfies zero energy and zero stress (strain-derivative of $W_1$) at zero strain, i.e., $\varepsilon=0$.

We model the constituent potential $W_2$ as 
\be
\begin{aligned}
    W_2(\varepsilon,\bfTheta) = & \underbrace{\mathcal{P}_{\omega_2}(\varepsilon-b,\bfTheta)}_\text{second convex potential} 
    -  \underbrace{\left(\mathcal{P}_{\omega_2}(0,\bfTheta) - v - W_1(b,\bfTheta)\right)}_\text{energy correction}\\
    &- \underbrace{(\varepsilon-b)\frac{\partial \mathcal{P}_{\omega_2}(0,\bfTheta)}{\partial \varepsilon}}_\text{stress correction},\\
    & \text{with}\qquad b = \calF_\beta(\bfTheta) \geq 0 \qquad \text{and} \qquad v = \calF_\nu(\bfTheta) \geq 0.
\end{aligned}
\ee
Here, $\mathcal{P}_{\omega_2}$ denotes another PICNN that is convex in $\varepsilon$ and contains trainable parameters $\omega_2$. However, unlike $W_1$, the energy and stress correction terms ensure that minimizer and minimum of $W_2$ are non-zero, i.e., $\varepsilon = b\geq 0$ and $W_2(b,\bfTheta)=v+W_1(b,\bfTheta)$. Both $b$ and $v$ act as offsets of $W_2$ with respect to $W_1$ in the strain~vs.~energy space. Their values are given by two additional classical MLP neural networks $\calF_\beta(\bfTheta)$ and $\calF_\nu(\bfTheta)$, parameterized by $\beta$ and $\nu$, respectively. The non-negativity constrain on $b$ and $v$ ensure that the non-convex combination of $W_1$ and $W_2$ in Eq.~\eqref{eq:Wrelax} yields a $W$ that is monotonically increasing (see \figurename~\ref{fig:FWD_Framework}a). {Note that, while we limit the construction of $W$ to a combination of only two convex potentials, additional potentials can be incorporated to model more complex nonlinear behavior if needed.}

We represent the experimentally generated dataset as 
\be
\calD=\left\{\left(\bfTheta^{(n)},\tilde\epsilon_t^{(n,i)},\tilde\sigma_t^{(n,i)}\right)\ :\  t=1,\dots,T{(n,i)};\ i=1,2,3;\ n=1,\dots,N\right\},
\ee
where $n$, $i$, and $t$ denote the different sample, direction of loading, and loadstep during loading, respectively. The total number of loadsteps $T(n,i)$ as well as change in strain between two loadsteps may not necessarily be the same across different samples and direction of loading. The forward ML model is then trained to minimize the mean absolute percentage error (MAPE) loss in stress predictions across the small training dataset:
\be
\omega_1,\omega_2,\beta,\nu\ \leftarrow\ \argminA_{\omega_1,\omega_2,\beta,\nu}\ \frac{1}{|\calD|} \ \sum_{n=1}^{N} \ \sum_{i=1}^{3}\ \sum_{t=1}^{T(n,i)}\ \left|\frac{\sigma\left(\tilde\varepsilon_t^{(n,i)},\bfTheta^{(n)}\right) - \tilde{\sigma}^{(n,i)}_t}{\tilde{\sigma}^{(n,i)}_t}\right|.
\ee
The loss function computes the relative error of the predicted and target values, ensuring that the predictions are forced to be accurate, regardless of the magnitude of the target value. Supplementary Note 3 presents the detailed ML training protocols.

Due to the limited amount of data, we take additional measures to facilitate the training process. For each data point, we distinguish between the type of spinodal topologies (lamellar, columnar, or cubic) and the loading directions ($\hat{\bfe}_1,\hat{\bfe}_2,$ and $\hat{\bfe}_3$). For each of these cases, we train distinct ML models and post-hoc lump them into a unified model; see Supplementary Note 3 for details.
\begin{figure}[t!]
	\includegraphics[width = 1.0\textwidth]{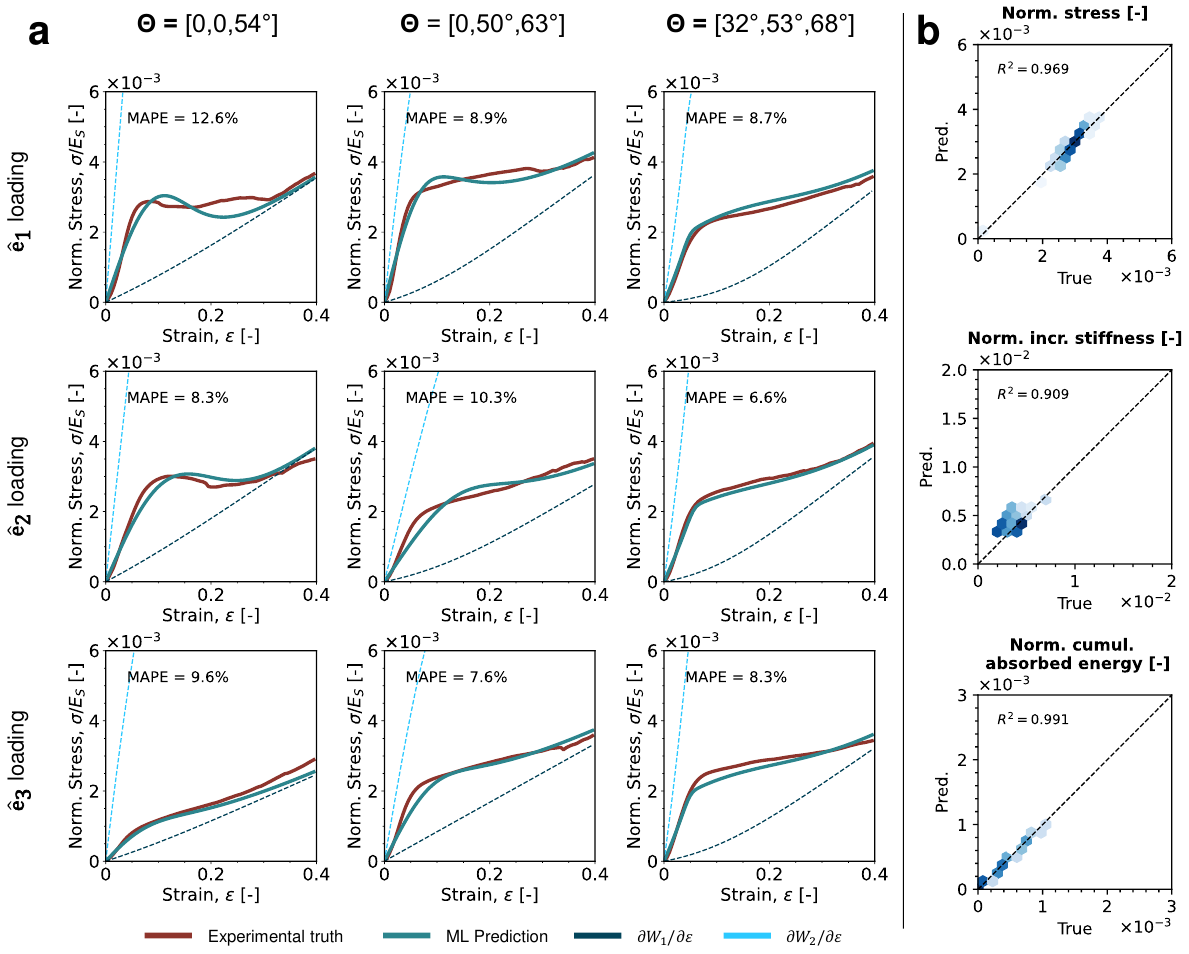}
	\centering
	\caption{Forward model results. (\textbf{a}) Stress-strain plots showing the ML-predicted (teal) versus experimental ground truth (red) curves for three representative spinodal design in the test dataset (i.e., sample from outside the training dataset). For reference, we show the derivatives $\partial W_1/\partial\varepsilon$ (dark blue) and $\partial W_2/\partial\varepsilon$ (light blue). (\textbf{b}) Distribution of predicted vs.~ground truth values on the test dataset for normalized stress, normalized incremental stiffness (i.e., slope of curve), and normalized cumulative energy absorbed (i.e., area under curve) at all strain increments. The dashed line represents the ideal line with zero intercept and unit slop; $R^2$ denotes the corresponding goodness-of-fit.} 
	\label{fig:FWD_results}
\end{figure}

We evaluate the predictive capabilities of our forward model by using test samples and their stress-strain data which the ML framework has not seen during training. In \figurename \ \ref{fig:FWD_results}a, we present the predicted (teal) and the experimentally measured (red) stress-strain curves across three directions each for three test samples---one each of lamellar, columnar, and cubic topologies. We also plot the derivatives ${\partial W_1}/{\partial \varepsilon}$ (dark blue) and ${\partial W_2}/{\partial \varepsilon}$ (light blue) of the two convex constituent energy potentials for reference. We generally observe a MAPE of 6--12\% across all types of spinodal structures and across all directions.
However, we do note that the errors for predicting the material behavior for lamellar structures (with small cone angles) are generally higher than for columnar and cubic structures. We attribute the higher errors for lamellar topologies to the relatively high sensitivity to imperfections and unpredictable localized mechanical behavior. \figurename \ \ref{fig:FWD_results}b illustrates the accuracy of the models across all the nine test cases. Despite the small dataset, we observe a goodness-of-fit $R^2>0.96$ for stress, $R^2>0.99$ for  absorbed energy  (i.e., cumulative area under the curve), and $R^2>0.90$ for the incremental stiffness (i.e., slope of the curve) with respect to the experimental ground truth at every strain point. Supplementary Note 2 provides details on how the absorbed energy  and incremental stiffness are computed. The lower accuracy in incremental stiffness predictions may be attributed to compounding of errors when computing derivatives of the stress-strain curve.

\subsection{Morphology-Dependent Deformation Mechanisms}
\begin{figure}[t!]
    \centering
    \includegraphics[width = 1.\textwidth]{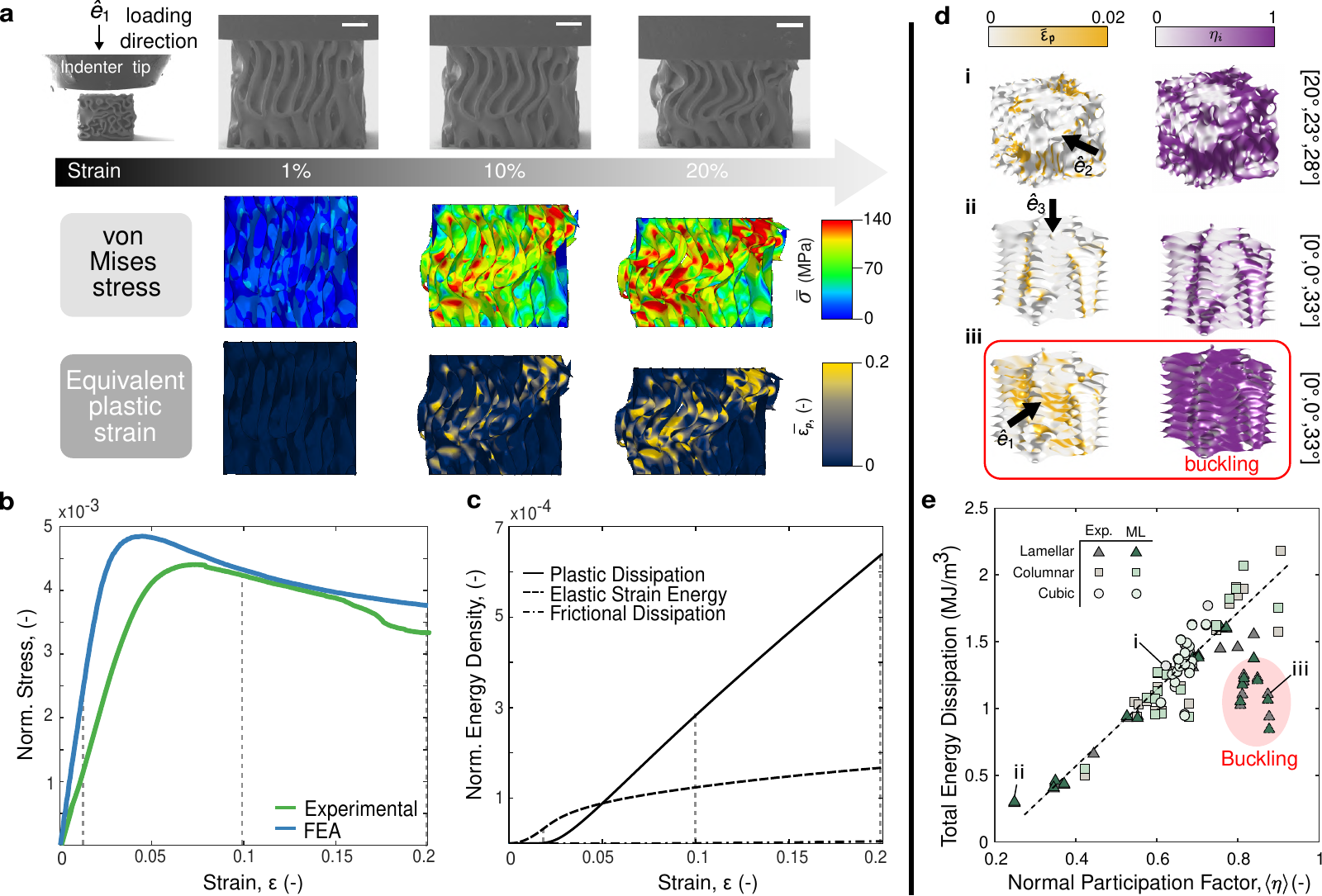}
    \caption{Finite-strain simulations and normal participation factor. (\textbf{a}) \emph{In situ} snapshots for $\bfTheta=[0^\circ,23^\circ,37^\circ]$ loaded in the $\hat{\bfe}_1$ direction for $1\%$, $10\%$, and $20\%$ strain points. Scale bars, 10 \textmu{}m. (\textbf{b}) Comparison between \emph{in situ} experiment (green line) and simulation (blue line) for the normalized stress-strain curves up to 20\% strain. (\textbf{c}) Distribution of energy mechanisms as a function of strain, including  plastic dissipation, elastic strain energy, and frictional dissipation. Up to 20 \% strain, friction effects are negligible and plastic dissipation is the dominant mode of dissipation. (\textbf{d}) Geometric representations of the normal participation factor $\eta$ (NPF) for three representative cases and its correlation to the equivalent plastic strain $\bar{\varepsilon}_p$. The dark purple regions correspond to $\eta \approx 1$  which correlates to regions that undergo high plastic deformation (gold regions) for the cases shown in (\textbf{\textit{i}}) and (\textbf{\textit{ii}}), while diminished correlation occurs in the buckling-prone case in (\textbf{\textit{iii}}). (\textbf{e}) Linear correlation between total energy dissipation and $\eta$,  where the data points corresponding to the three cases in (\textbf{d}) are indicated, showing a loss of correlation for geometries that undergo buckling events.}
    \label{fig:FEA}
\end{figure}

To bridge the gap between the measured/predicted nonlinear responses and the morphology descriptors that lead to various deformation mechanisms, we employed nonlinear finite element models of selected morphologies within our design space. We leveraged these simulations to add mechanistic insight to performance metrics such as absorbed energy and its relation to our spinodal design space, particularly linking nonlinear mechanisms to surface-curvature distributions in a given morphology. As representative designs of the lamellar, columnar, and cubic categories, we selected the $\bfTheta=[0^\circ,0^\circ,33^\circ]$, $\bfTheta=[0^\circ,23^\circ,37^\circ]$, and $\bfTheta=[20^\circ,23^\circ,24^\circ]$ morphologies, respectively, to model within the finite element framework.  

To match the conditions in our experiments, we modeled each morphology with no constrains on the lateral faces while the degrees of freedom of the bottom nodes were fully constrained and a compressive displacement was imposed on the top nodes. To capture complex responses while managing the computational cost of each simulation, we discretized each morphology using structural shell elements endowed with an elasto-plastic material model, where the plastic flow curve was determined from experiments on the constituent polymer (see Supplementary Note 2). {We note that this plasticity model was chosen for its simple implementation, while it is not intended to fully capture the physical mechanisms of permanent deformation in our polymer material. As a result, we employ these simulations as a tool to uncover differences in deformation mechanisms between representative morphologies, instead of as an all-encompassing material model for our spinodal samples. Improvements to the model can be made to account for polymer-specific physics during nonlinear deformation \cite{ANANDamorpsolidsmodel,damagemodel,saelen2023stereolithModel}, but are beyond the scope of this work.}  

Comparing the simulations to the \emph{in situ} experimental responses demonstrated kinematic agreement within all morphologies, prior to the emergence of fracture events within the polymeric shells which were not intended to be captured numerically. For instance, simulations of compression on the $\bfTheta=[0^\circ,23^\circ,37^\circ]$  morphology along the $\hat{\bfe}_1$ direction (\figurename~\ref{fig:FEA}a, other morphologies/directions in Figs. S13 and S14) captured nonlinear responses such as buckling and self-contact. In agreement with experiments, the simulations predicted the buckling events to emerge at $\varepsilon \approx 5\%$, accompanied by stress localization at regions of higher curvature. 
Beyond the onset of buckling, the simulations predicted accumulation of equivalent plastic strain $\bar{\varepsilon}^p$ primarily in these high-curvature regions, agreeing with the sites of fracture initiation in \emph{in situ} experiments.
While the linear response of the simulations proved to be consistently stiffer than that observed in experiments (\figurename~\ref{fig:FEA}b), as expected in a model that is agnostic of minor fabrication defects, the qualitative large-deformation response was in agreement and the effective yield strength $\sigma_y^*$ predictions were within 2\% of experimental values---with the exception of the lamellar sample whose instability-driven onset of nonlinearity was within 13\%. 
Most importantly, the simulations provided estimates of energy distributions within the probed morphologies, suggesting energy absorption to be primarily due to plastic dissipation. Specifically, plastic dissipation was calculated to equal the elastic strain energy in the structure at compressive strains of 5\%, shortly after the emergence of buckling (\figurename~\ref{fig:FEA}c). Even at strains of 20\% in various morphologies, contributions due to frictional dissipation were minimal, with plastic dissipation accounting for up to 78\% of the total internal energy in the system in the case of the lamellar architecture. 
 
 To establish a connection between these local phenomena and the underlying geometry, we introduce a parameter $\eta$ termed the normal participation factor (NPF), which denotes the degree to which surfaces within an architecture align with the loading direction. For a given surface element $i$, the NPF  $0 \leq \eta_i \leq 1$ is computed as
\be
    \eta_i=1-(\mathbf{n_i}\cdot\mathbf{e_d})^2,
\ee
where $\mathbf{n_i}$ represents the normal vector of the element and $\mathbf{e_d}$ denotes the direction of loading. The NPF assumes a value of 1 when the loading direction is parallel to the surface, while it approaches 0 when the loading direction is perpendicular. Using these finite element models to determine plastic strain localization in spinodal morphologies reveals a correlation between the regions of high NPF and localized equivalent plastic strain for a majority of the morphologies (\figurename~\ref{fig:FEA}d (\textbf{\textit{i}, \textit{ii}})).   
However, some morphologies such as  $\bfTheta=[0^\circ,0^\circ,33^\circ]$ under loading in the $\hat{\bfe}_1$ direction  show a greatly diminished correlation between the distribution of $\bar{\varepsilon}_p$ and NPF values (\figurename~\ref{fig:FEA}d (\textbf{\textit{iii}})). This discrepancy arises due to buckling phenomena, which lead to less efficient load distribution upon buckling and corresponding localized strain accumulation. 
Using these local $\eta_i$ values, we define the average NPF for the entire morphology under a given loading direction, denoted as $\langle\eta\rangle$, as
\be
\langle\eta\rangle=\frac{\sum_{i=1}^NA_i\eta_i}{\sum_{i=1}^NA_i},
\ee
where $A_i$ represents the area of the $i$\textsuperscript{th} surface element. Plotting the $\langle\eta\rangle$ against the total energy dissipation, as shown in \figurename~\ref{fig:FEA}e, reveals a notable correlation between the two parameters. Points deviating from the trend line, signify exceptional cases characterized by buckling instabilities. Furthermore, the predictions obtained from the ML algorithm are overlaid, demonstrating its ability to capture the large-deformation effect of these instabilities. These exceptional cases underscore the necessity for employing two potentials within the ML framework, rather than one, to accurately represent the nonlinear energy absorption associated with buckling phenomena in spinodal metamaterials. 

\subsection{Inverse Design for Tailored Stress-Strain Response}
\begin{figure}[b!]
	\includegraphics[width = 1.0\textwidth]{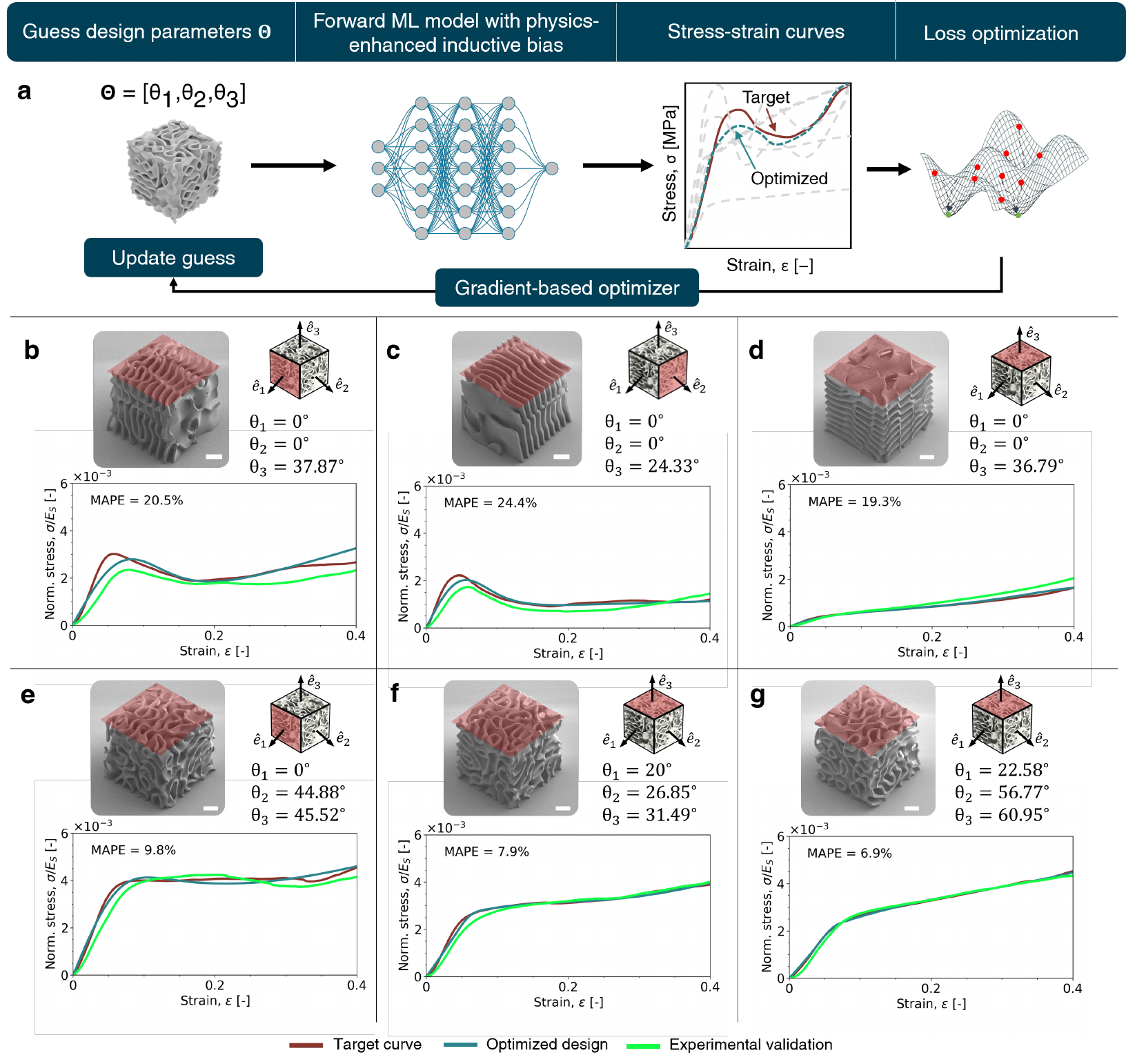}
	\centering
	\caption{Inverse design framework and results. (\textbf{a}) The forward ML model serves as a fast surrogate to evaluate the structure-property relations within an iterative gradient-based optimization scheme. The input to the optimization scheme is a target stress-strain curve (indicated in dark red) and the optimizer iteratively updates the guess design parameters until the loss between the predicted and target stress-strain curve is minimized. (\textbf{b}-\textbf{g}) We optimize for six target stress-strain curves which are not present in the training dataset. The stress-strain plots show the target (red), optimized (teal), and experimentally verified (green) stress-strain curves (with MAPEs specified between the target and experimental response). For each optimized design, we provide SEM images of the fabricated samples corresponding to the optimized design parameters and the green stress-strain curves, with the vertical direction (substrate normal direction) corresponding to the loading direction highlighted in transparent red. A graphic of the standard axis convention used in this work is given for reference to loading direction. Scale bar, 20 \textmu{}m.}
	\label{fig:OPT_Framework}
\end{figure}

We use the predictive capabilities of the forward ML model as a fast  surrogate (to experiments) for inverse designing spinodal metamaterials with prescribed nonlinear stress-strain responses. Let $\calS=\{(\hat\varepsilon_t,\hat\sigma_t)\ : \ t=1,\dots,T\}$ be a desired stress-strain response for quasi-static uniaxial compression loading discretized on $T>0$ points. We use the MAPE loss to formulate the  inverse design task as an optimization:
\be\label{eq:opt}
\bfTheta, \hat\bfe_i\leftarrow \argminA_{\bfTheta,\ i\in\{1,2,3\}} \  \frac{1}{T}\sum_{t=1}^T \left|\frac{\sigma^{(i)}(\hat\varepsilon_t,\bfTheta) - \hat\sigma_t}{\hat\sigma_t}\right|.
\ee
The stress predictions $\sigma^{(i)}(\hat\varepsilon_t,\bfTheta)$ are evaluated via the forward ML model using Eq.~\eqref{eq:sigma}, where the superscript $(\cdot)^{(i)}$ indicates the model corresponding to the $\hat\bfe_i$ principal direction of loading (see Supplementary Note 4).
This has two advantages: \textit{(i)} the numerical optimization requires several evaluations of the stress predictions, which are orders of magnitude faster when performed via the forward ML model than experiments (or even simulations, if feasible); and \textit{(ii)}  while gradient-based optimization schemes (e.g., gradient descent \cite{hardt2016train}) are more efficient and stable than non-gradient-based methods, they require computing the sensitivities of the loss function with respect to the variables (in this case, $\bfTheta$). Leveraging the computational graph and backpropagation of NNs \cite{LeCun2015}, the forward ML model enables automatic differentiation \cite{Paszke2019} of the stress predictions $\sigma$ and in turn, the loss function with respect to $\bfTheta$. In contrast to computationally intensive numerical differentiation (perturbing the input and re-evaluating the loss function), automatic differentiation provides analytically exact gradients and enables a more stable numerical optimization. 

Here, we use the Adaptive Moment Estimation or Adam \cite{kingma2014adam} optimizer to solve Eq.~\eqref{eq:opt}. We highlight that the inverse design challenge is ill-posed as multiple designs can exhibit the target stress-strain curve. To bypass this challenge, we perform the optimization for different initial guesses in parallel and select the design with the lowest MAPE loss at the end. Supplementry Note 4 provides additional details on the optimization protocols and a pseudocode is presented in Supporting Information Algorithm 1.

To demonstrate the efficacy of our framework, we inversely design for target curves from our dataset, which were multiplied by a factor $\kappa=1.2$. This ensures that the target mechanical behavior is beyond the stress-strain curves which were provided in the limited experimentally generated training set. We present six distinct target curves and their respective optimization results in \figurename  \ \ref{fig:OPT_Framework}b-g. For each target (red) we optimize the design parameters to obtain an optimized curve (teal). Finally, we fabricate and test the structures with the optimized design parameters (green). The first target (b) has a strain-softening region with a subsequent strain-hardening region.  Similarly, the second target (c) has a strain-softening region after the yield point, with a subsequent stress plateau. The third target (d) exhibits a relatively stable, monotonically increasing stress-strain response. The fourth target (e) has a relatively stiff linear elastic region and a subsequent stress plateau. The last two targets (f-g) possess stiff linear elastic region with a subsequent strain-hardening behavior.
For the first target, the optimization framework proposes a lamellar structure ($\bfTheta=[0^\circ,0^\circ,37.87^\circ]$) oriented along the $\hat{\bfe}_1$-direction to achieve the desired stress-strain response. The fabricated structure matches the target response well with a MAPE of $20.5\%$. Most of the error stems from the strain-hardening region in the fabricated sample, which the forward module overpredicted. For the second target, also a lamellar structure ($\bfTheta=[0^\circ,0^\circ,24.33^\circ]$) oriented along the $\hat{\bfe}_2$-direction, the MAPE between the response of the fabricated sample and the target curve is $24.4\%$. Here, the three curves are offset by a small margin, with the experiment showing a lower yield stress than what was predicted. The overall macro-structural behavior, i.e., strain-softening with subsequent stress-plateau, is captured with the optimized design parameters. For the third target, the fabricated target ($\bfTheta=[0^\circ,0^\circ,36.97^\circ]$) exhibits a higher strain-hardening than what was predicted, which results in an MAPE of $19.3\%$. However, similarly to the second target, the overall stress-strain behavior is well captured. For the fourth, fifth, and sixth target, we observe excellent matches with the queried stress-strain behaviors with MAPEs of $9.8\%, 7.9\%,$ and $6.9\%$.

Validation on these six target curves demonstrates the validity of the proposed framework to capture highly nonlinear responses in complex 3D spinodal morphologies that lack internal symmetry and periodicity. While quantitative agreement between experimental and optimized responses improves for columnar or cubic morphologies, which have less pronounced anisotropy compared to lamellar ones, the overall qualitative response was always captured regardless of the $\bfTheta$ representation. Specifically, in scenarios where micromechanical instabilities or failure events led to a macro-structural negative-stiffness responses, the framework accurately accounted for these features with closely matched stress and strain levels. Moreover, our results demonstrate that the physical basis chosen for our ML framework accurately describes the structural response of thick-shell spinodal structures at finite strains---a complex parameter space that is intractable to fully explore through high-fidelity simulations or experiments alone.    

\section{Conclusion}
Designing complex spinodal metamaterials with a wide range of topologies and corresponding nonlinear mechanical behavior is challenging, especially when computational modeling can be costly and experimental data is scarce. Here, we introduced a physics-enhanced ML and optimization framework that bypasses this challenge by directly using extremely sparse experimental data and enables the inverse design of spinodal structures with tailored finite-strain mechanical responses. 

The lack of data for learning the structure-property relations is compensated by the physics-based inductive biases, which aid in identifying nonlinear responses such as instability- and localization-dominated responses. Tracing these responses back to nonconvex energetic potentials allows for a versatile framework that may be applied to a variety of lightweight microstructures employed in the mechanical metamaterials community. Inspired from phase transformation modeling approaches, combining multiple convex (in strain) neural networks to form nonconvex but monotonically increasing potentials can accurately and efficiently capture complex nonlinear stress-strain behavior in presence of extreme and localized deformation including failure. At the same time, partial input convexity of PICNN architectures allows capturing arbitrary non-convex functional relations with the design parameters of spinodal metamaterials.

In this work, we verify the importance of the above described physical basis in the ML model by looking at the local deformation mechanisms. Despite the computational cost, finite element simulations lead to the observation that plastic dissipation is the dominant dissipation mechanism up to $20\%$ strain and thus the geometric normal participation factor $\eta$ was introduced to describe the local loading distributions. We found that $\eta$ and energy absorption of our samples, were strongly correlated. However, the samples that underwent buckling had diminished correlation due to premature strain localization, evidencing the physics-enhanced ML as a route to accurately and efficiently predict energy absorption for these unique cases and highlighting the need for the second convex energy well.     

While we developed this framework for spinodal metamaterials, we note that the ML inverse design framework and the physics-enhanced inductive biases are sufficiently general to be  individually (or in combination) adapted for any class of metamaterials (e.g., truss, plate, or TPMS lattices) and data representation (e.g., vector, graph, or pixel/voxel-based parameterizations). The gradient-based optimization strategy  can also be adapted to other approaches, including generative ML methods such as conditional variational autoencoders \cite{glaesener2023predicting} and diffusion models \cite{bastek2023inverse}. {Additionally, one could extend the model to handle multiple loading cycles by introducing internal parameters that track the material’s loading history. These additions could allow the model to distinguish between loading and unloading phases and account for plastic deformation.}
Moreover, in this work, we have performed a comparatively large number of high-fidelity experiments in both \emph{ex situ} and \emph{in situ} formats to assist in the understanding of the ML model from a physical basis. 
We leverage microscale fabrication and testing to obtain an order-of-magnitude increase in data throughput---necessary for applicability to ML frameworks. From the 321 experiments performed in this work, we were able to capture the complex localized deformation behavior of thick-shelled spinodal morphologies, which led us to gain critical insights into the physical interpretation of our ML framework results. This work closes the gap in understanding the structure-to-property relations of high-relative-density spinodal morphologies at finite strains, relevant to applications of high energy absorption materials. The combination of using an experimentally generated dataset to train a physics-enhanced ML framework shows a promising avenue for designing and understanding the complex architected materials of the future.

\section*{Methods}

Details on the virtual specimen generation (Supplementary Note 1), the sample fabrication, physical characterization and finite element analysis (Supplementary Note 2), the data preprocessing, the PICNN-framework and training protocols (Supplementary Note 3), and the gradient-based multi-initialization optimization scheme (Supplementary Note 4) are provided in the Supplementary information.

\clearpage
 \bibliographystyle{elsarticle-num} 
 \bibliography{Bib}

\section*{CRediT authorship contribution statement}

\textbf{Prakash Thakolkaran:} Formal analysis, Data curation, Investigation, Methodology, Software, Visualization, Writing – original draft, Writing – review and editing. \textbf{Michael Espinal:} Formal analysis, Data curation, Investigation, Visualization, Validation, Writing – original draft, Writing – review and editing. \textbf{Somayajulu Dhulipala:} Formal analysis, Data curation, Writing – original draft, Writing – review and editing. \textbf{Siddhant Kumar:} Conceptualization, Funding acquisition, Resources, Supervision, Writing – review and editing. \textbf{Carlos M. Portela:} Conceptualization, Funding acquisition, Resources, Supervision, Writing – review and editing.

\section*{Data availability}
The data and codes generated in the current study are freely available at \url{https://github.com/mmc-group/finite-strain-inverse-designed-spinodoids}.
\section*{Declaration of Competing Interest}
The authors declare no competing interests. 
\section*{Acknowledgements}

Carlos M. Portela acknowledges financial support from the National Science Foundation (NSF) CAREER Award (CMMI-2142460), and the MIT MechE MathWorks Seed Fund and Engineering Fellowship Program. This work was carried out in part through the use of MIT.nano’s facilities. This material is based upon work supported by the National Science Foundation Graduate Research Fellowship Program under Grant No. 2141064. Any opinions, findings, and conclusions or recommendations expressed in this material are those of the author(s) and do not necessarily reflect the views of the National Science Foundation.

\end{document}


\begin{center}
    \section*{{\textbf{Supplementary Information}}}

   \section*{\textbf{Experiment-Informed Finite-Strain Inverse Design
of Spinodal Metamaterials}}
\end{center}
Prakash Thakolkaran\textsuperscript{1,$\dagger$}, Michael Espinal\textsuperscript{2,$\dagger$}, Somayajulu Dhulipala\textsuperscript{2}, \\Siddhant Kumar\textsuperscript{1,*,$\ddagger$}, Carlos M. Portela\textsuperscript{2,*,$\ddagger$}
\\
\section*{Affiliations}
\textsuperscript{1}Delft University of Technology, Department of Materials Science and Engineering,  2628 CD Delft, The Netherlands
\\
\textsuperscript{2}Massachusetts Institute of Technology, Department of Mechanical Engineering, 77 Massachusetts Avenue, Cambridge, MA 02139, USA\\
\textsuperscript{*}cportela@mit.edu, sid.kumar@tudelft.nl\\
\textsuperscript{$\dagger$,$\ddagger$}these authors contributed equally to this work

\section*{This PDF file includes:}
Supplementary Note 1. Virtual specimen generation \\
Supplementary Note 2. Fabrication, experiments, and finite element analysis \\
Supplementary Note 3. Machine learning framework \\
Supplementary Note 4. Optimization framework details\\
Supplementary Figures S1-S14.\\
Supplementary Tables S1-S4.\\
Algorithm 1.\\
Supplementary References. 

\subsection*{Other Supplementary Material}
Supplementary Movies S1-S9.
\pagebreak

\section*{Supplementary Note 1. Virtual specimen generation}
We summarize all the parameters used in the data generation process in Supplementary Table \ref{tab:datagen_parameters}. Each microstructure is represented by the three cone angles $\bfTheta=\{\theta_1,\theta_2,\theta_3\}$. The angles are always chosen in ascending order, i.e., $\theta_1 \leq \theta_2 \leq \theta_3$ to reduce the design complexity, while noting that the permutation symmetry allows the stress-strain curves for other parameters to be automatically inferred (e.g., the response along $\hat{\bfe}_1$ for $[\theta_1,\theta_2,\theta_3]$ is equivalent to the response along $\hat{\bfe}_2$ for $[\theta_2,\theta_1,\theta_3]$). 
The rest of the parameters are chosen as constants.

\begin{table}[h]
\centering
\caption{List of parameters used for the data generation.} 
\label{tab:datagen_parameters}
\begin{tabular}{lcc}
\hline
\multicolumn{1}{l}{\textbf{Parameter}} & \textbf{Notation} & \textbf{Value} \\ \hline
Resolution of SDF &$K$ &  $100$ \\
RVE size &$\ell$ &  100 \textmu{}m \\
Wave number & $\beta$ & $5/\ell$ \\
Surface thickness & $h$ & $3\mu$m \\
SDF radial smoothing coefficient &$\lambda_r$& $0.3$ \\
SDF angular smoothing coefficient &$\lambda_{\phi}$& $0.175$ \\
Lower bound cone angle &$\theta_{\min}$& $20^{\circ}$ \\
Upper bound cone angle &$\theta_{\max}$& $70^{\circ}$ \\
Energy interpolation constant & $k_T$ & $0.7$ \\
Number of lamellar data points & $-$ & $11$ \\
Number of columnar data points & $-$ & $36$ \\
Number of cubic data points & $-$ & $60$ \\
Total number of data points & $N$ & $107$ \\
Number of interpolated strain steps &$T$ &  $80$ \\
First strain step &$\varepsilon_1$ &  $0.001$ \\
Last strain step &$\varepsilon_T$ &  $0.4$ \\
Uniform strain interval &$\Delta \varepsilon$ &  $(\varepsilon_T-\varepsilon_1)/(T-1)$ \\\hline
\end{tabular}
\end{table}

The cone angles---when non-zero---are sampled as integers in the range of $[\theta_{\min},\theta_{\max}]$. We used the dataset of Kumar et al.\cite{kumar2020inverse} as a basis for sampling the design parameters. We randomly sampled a total of $107$ spinodoid parameters $\bfTheta$, with $11$ lamellar, $36$ columnar, and $60$ cubic structures. These 107 samples, along with their characterizations in three directions, result in 321 analyzed samples, corresponding to 609 effective $\bfTheta$ parametrizations. This can be calculated as 11 unique lamellar structures, each with 3!/2! permutations (since two of the angles are zeros and only one unique angle is present), and 96 columnar and cubic samples, which have 3! permutations (since all the cone-angles are unique). Thus, the total is given by 11$\times$3!/2! + 96$\times$3! = 609.

\begin{figure}[t]
	\includegraphics[width = 1.00\textwidth]{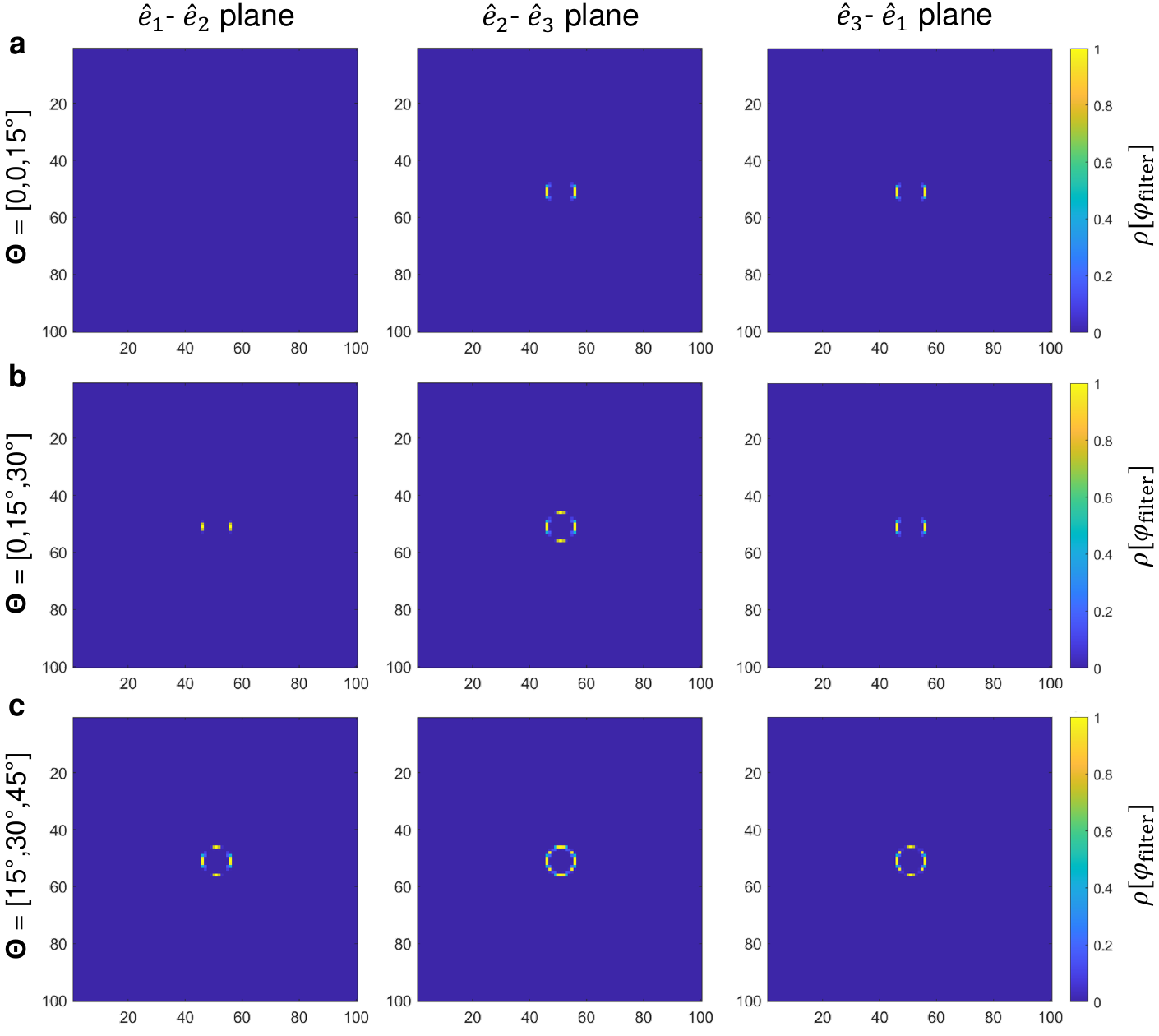}
	\centering
	\caption{Planar projections of the SDF $\rho[\varphi_{\text{filter}}]$ for a representative (a) lamellar, (b) columnar, and (c) cubic structure.}
	\label{fig:FFT}
\end{figure}

Following the sampling of the design parameters $\bfTheta$, we generate the structures using the Fourier-based approach described in the main text. \figurename \ \ref{fig:FFT} shows three sample slices of the SDF $\rho[\varphi_{\text{filter}}]$ for representative (a) lamellar, (b) columnar, and (c) cubic structures. The three columns correspond to the $\hat{\bfe}_1$-$\hat{\bfe}_2$, $\hat{\bfe}_2$-$\hat{\bfe}_3$, and $\hat{\bfe}_3$-$\hat{\bfe}_1$ plane. In \figurename \ \ref{fig:FFT}a we observe that in the middle of the $\hat{\bfe}_1$-$\hat{\bfe}_2$ plane all values in the SDF are zero, since the cone is aligned with the $\hat{\bfe}_3$-axis. However, in the $\hat{\bfe}_2$-$\hat{\bfe}_3$, and $\hat{\bfe}_3$-$\hat{\bfe}_1$ plane, the cone outline (at distance $\beta$ from the origin) is clearly indicated with $\theta_3=30^\circ$. Similarly for (b), we  see the outlines of two cones -- one aligned with the $\hat{\bfe}_3$-axis with $\theta_3=30^\circ$, and one aligned with the $\hat{\bfe}_2$-axis with $\theta_2=15^\circ$. Lastly, for the cubic case in (c), we  see all the three cones, namely, the cone aligned with the $\hat{\bfe}_3$-axis ($\theta_3=45^\circ$), the one aligned with the $\hat{\bfe}_2$-axis ($\theta_2=30^\circ$), and the one aligned with the $\hat{\bfe}_1$-axis ($\theta_1=15^\circ$). All the SDFs were produced with the smoothing operation, which is why we observe the smear along the radial and angular directions. 
Following the morphology generation using the Fourier approach, we generate the STLs of the zero-isosurfaces in MATLAB, and followed by a thickening procedure described in the materials and methods. 

\section*{Supplementary Note 2. Fabrication, experiments, and finite element analysis}

\subsection*{Selection of thickness and wavenumber}

\begin{figure}[ht]
	\includegraphics[width = 1.0\textwidth]{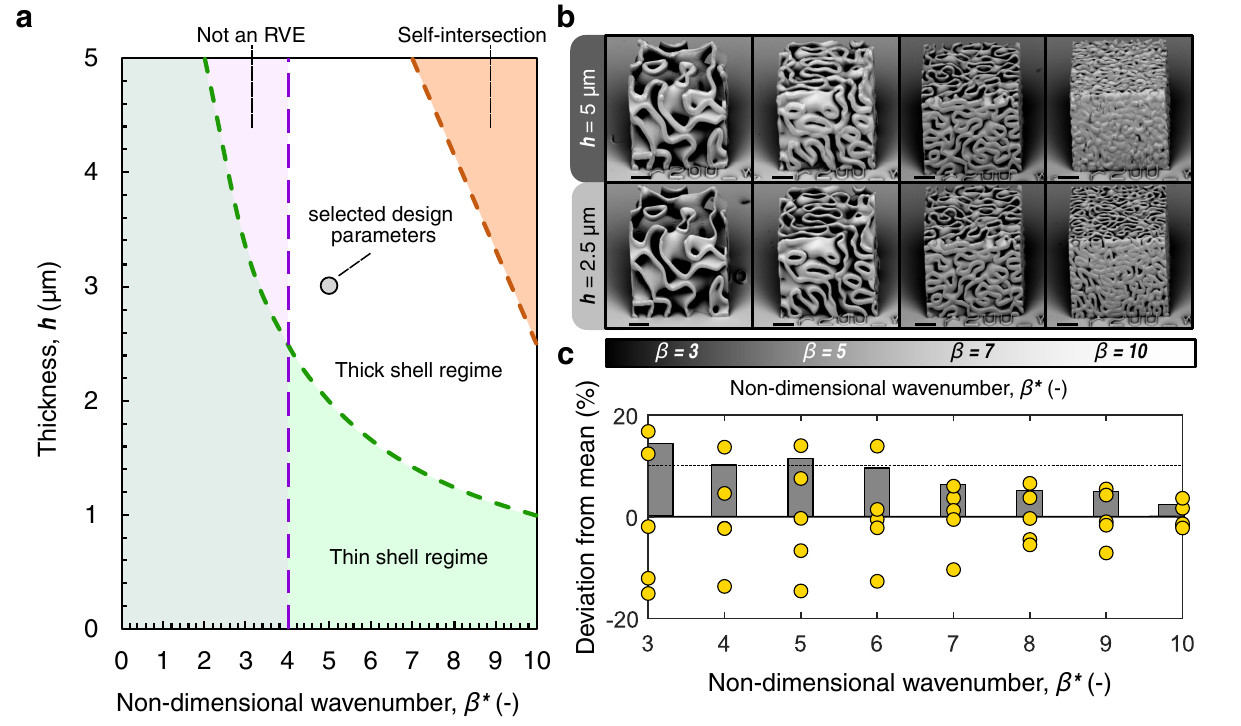}
	\centering
	\caption{(a)  Plot of thickness versus wavenumber demarcating the regions of RVE convergence, thin shell regime and self-intersection. (b) SEM images of spinodoid morphologies fabricated with thickness of 2.5 \textmu m and 5 \textmu m for $\beta^*$ = 3,5,7 \& 10. (Scale bar = 20 \textmu m) (c) Deviation of stiffness from mean versus the wavenumber for 5 spinodoid morphologies generated per wavenumber. The bar represents the standard deviation and the dashed line represents 10\% error.} 
	\label{fig: RVEselection}
\end{figure}

Three considerations guide the selection of the wavenumber and thickness for spinodoid morphologies. Firstly, the wavenumber, representing the length scale of features, must be sufficiently high for the generated morphology to serve as a representative volume element (RVE). Secondly, the spinodoid morphology should possess adequate thickness for a given wavenumber to be considered a thick shell. Lastly, the fabrication of these morphologies should avoid self-intersection.

\figurename~\ref{fig: RVEselection}a presents a design plot illustrating the relation between thickness and non-dimensional wavenumber, $\beta^*=\beta \ell$, where $\ell$ is the unit cell size, with limits corresponding to the three design considerations. In \figurename~\ref{fig: RVEselection}b, SEM images showcase fabricated spinodoid morphologies across various thicknesses and non-dimensional wavenumbers. It is evident that, at a thickness of 2.5 \textmu m and $\beta^*=10$, self-intersection occurs. Even at a lower value of $\beta^*=7$, self-intersection is observed, particularly on the vertical faces for a thickness of 2.5 \textmu m.

To assess RVE convergence, we performed finite element simulations for five spinodoid morphologies generated for each non-dimensional wavenumber from $\beta^*=3$ to 10, with $\bfTheta = [45^\circ,45^\circ,45^\circ]$ on Abaqus. Static linear perturbation and free boundary conditions were applied to obtain stiffness in the $\mathbf{\hat{e}_3}$ direction, utilizing shell (S3) elements as the discretization in the simulations. The results, depicted in \figurename~\ref{fig: RVEselection}c, illustrate the deviation of obtained stiffness from the mean stiffness for each non-dimensional wavenumber. Notably, for $\beta^*=4$ and above, the standard deviation of the points is within a 10\% error range.

Finally, to ensure we are in the thick-shell regime, the thickness of the morphologies was selected such that the product of the curvature and the thickness is greater than 0.1\cite{jnreddyshells}. For a given non-dimensional wavenumber $\beta^*$, the curvature $\kappa$, is approximately given by $\kappa \approx \beta^*/l$. Therefore, to be in the thick shell regime

\begin{equation}
\kappa h=\frac{\beta^* h}{l}>0.1.
\end{equation}

The curve corresponding to $(\beta^*h)/l=0.1$ is shown in Fig.~\ref{fig: RVEselection}a, demarcating the thin- and thick-shell regimes. Our selected design corresponds to a nominal thickness of 3 \textmu m and a non-dimensional wavenumber, $\beta^*=5$, which satisfies all three design considerations.

\subsection*{Two-Photon Polymerization 3D Printing}
After the zero-isosurfaces are obtained for the 107 different spinodal parameters, $\bfTheta$, each STL is then thickened to the chosen nominal thickness of 3 \textmu m. The thickening procedure was done in Blender (3.3.1) using a solidify modifier. After the final parameters were selected, the morphologies were fabricated using a two-photon lithography process with a Nanoscribe Photonic Professional GT2 (Nanoscribe GmbH) system, using IP-Dip photoresist and a 63$\times$ objective. A scan speed of 10,000 \textmu{}m/s, laser power of 25 mW, slicing distance of 0.3 \textmu{}m, and hatching distance of 0.2 \textmu{}m were used. Post-printing, the samples were developed in propyleneglycol-monomethyl-ether-acetate (PGMEA) for 40 to 60 minutes followed by immersion into isopropyl alcohol to remove any remaining PGMEA. Lastly, the samples underwent a critical point drying process 
(Tousimis Autosamdri-931). The fabricated unit cells had a relative density of approximately 40\%, with average size of 92$\pm$1.1 $\times$ 92$\pm$1.5 $\times$ 92$\pm$1.1 \textmu{}m$^3$ and a shell thickness of 2.6$\pm$0.1 \textmu{}m. An example of the print quality is shown in \figurename~\ref{fig:Defeacts}.   

\subsection*{Physical Characterization}

\begin{figure}[ht]
	\includegraphics[width = 0.6\textwidth]{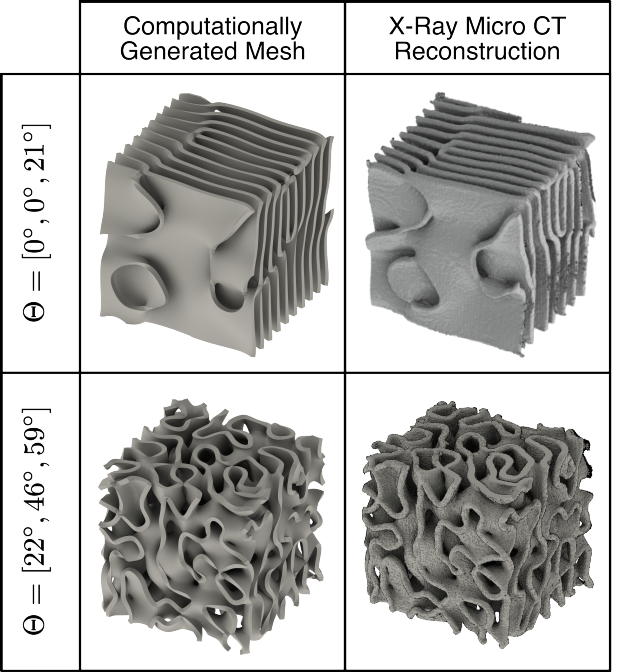}
	\centering
    	\caption{Computationally generated and X-Ray \textmu-CT 3D Reconstructions of $\Theta=[0^{\circ},0^{\circ},21^{\circ}]$ and $\Theta=[22^{\circ},46^{\circ},59^{\circ}]$ morphologies show great match and lack of discernible defects.} 
	\label{fig:xct}
\end{figure}

Representative spinodal morphologies were examined using X-ray \textmu-Computed Tomography (X-Ray \textmu{}CT). Three-dimensional reconstructions were conducted employing a marching-cube algorithm subsequent to binarization of the z-stack acquired from X-Ray \textmu{}CT. \figurename~\ref{fig:xct} displays the computationally generated morphology alongside the corresponding 3D reconstruction for the $\bfTheta=[0^{\circ},0^{\circ},21^{\circ}]$ and $\bfTheta=[22^{\circ},46^{\circ},59^{\circ}]$ morphologies. The nominal relative densities of the morphologies $\bfTheta=[0^{\circ},0^{\circ},21^{\circ}]$ and $\bfTheta=[22^{\circ},46^{\circ},59^{\circ}]$ were calculated as 31.4\% and 33.8\%, respectively, while the actual relative densities were determined to be 41.8\% and 35.8\%, respectively. The increase in relative density can be ascribed to shrinkage of the unit cell after polymerization and development and the larger-than-nominal thickness of the fabricated shells.

\subsection*{Material characterization}
\begin{figure}[ht]
	\includegraphics[width = 0.5\textwidth]{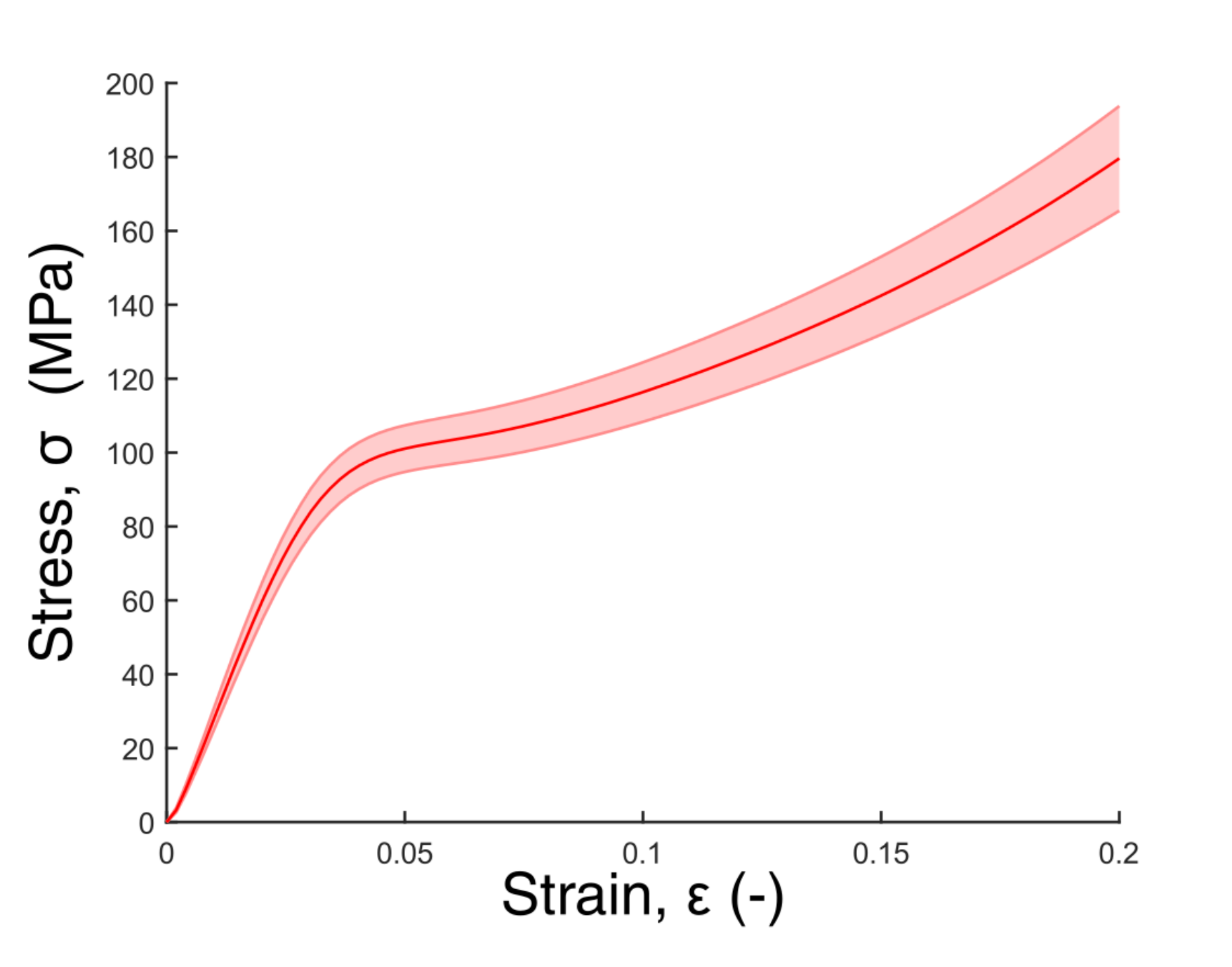}
	\centering
	\caption{Plot showing the mean (bold line) and standard deviation of 20 randomly sampled IP-Dip pillar compression up to 20 \% strain.} 
	\label{fig: Pillar conversion}
\end{figure}

For our experimental data to be applicable to polymeric spinodal metamaterials made of a different constituent material, we choose to normalize the stress by the IP-Dip Young's modulus when presenting our data. To ensure consistency across print batches we fabricated a minimum of six monolithic pillars along with each set of spinodal architectures (using the same print parameters), and performed uniaxial compression experiments to determine the stiffness and yield strength of the constituent polymer. The average diameter and height of all pillars were 27.5$\pm$0.5 \textmu{}m and 94.4$\pm$2.3 \textmu{}m, respectively. All pillars were compressed \emph{ex situ} and their stiffness was measured by applying a linear fit to the linear loading region of the stress-strain curve, while the yield strength was calculated using the 0.2\% offset method. \figurename~\ref{fig: Pillar conversion} shows 21 stress-strain curves from 6 batches. The modulus and yield strength for the IP-Dip photoresist were determined to be $3.2 \pm 0.3$ GPa and $77 \pm 7$ MPa, respectively.

\subsection*{Mechanical characterization}
\begin{figure}[ht]
	\includegraphics[width = 1.0\textwidth]{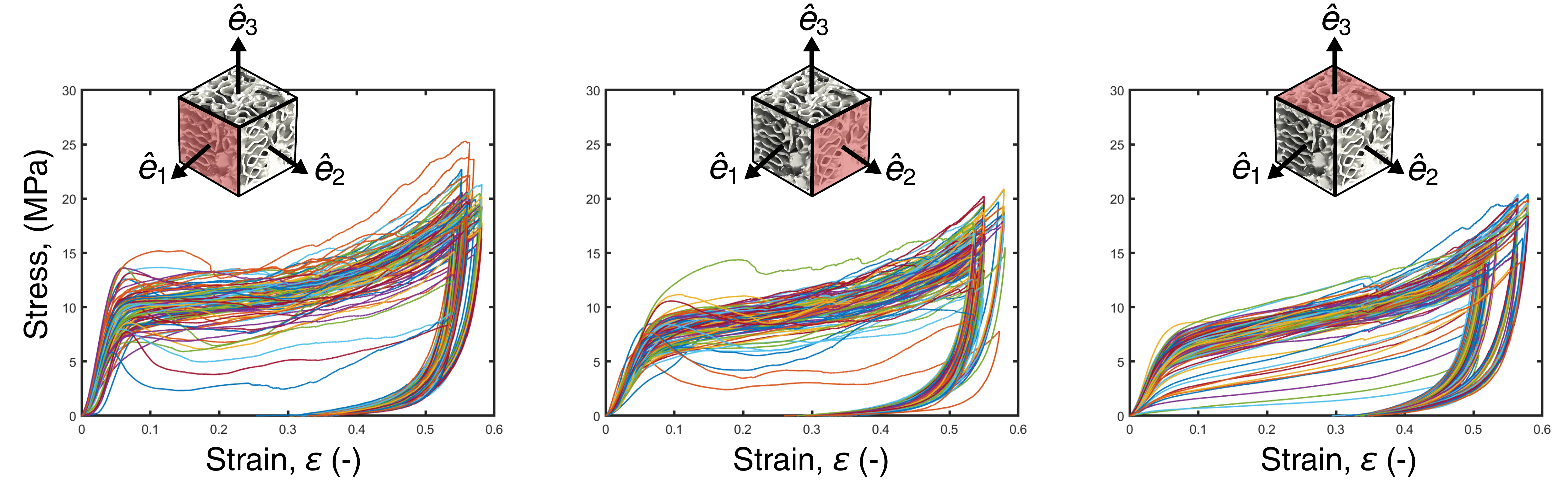}
	\centering
	\caption{The stress-strain plots for the entire dataset for loading in the  $\hat{\bfe}_1$ (left), $\hat{\bfe}_2$ (center), and $\hat{\bfe}_3$ (right) principal directions. The loading directions are highlighted in the schematic above the plots to visualize the loading direction.} 
	\label{fig:all_experiments}
\end{figure}

\begin{figure}[ht]
	\includegraphics[width = 0.75\textwidth]{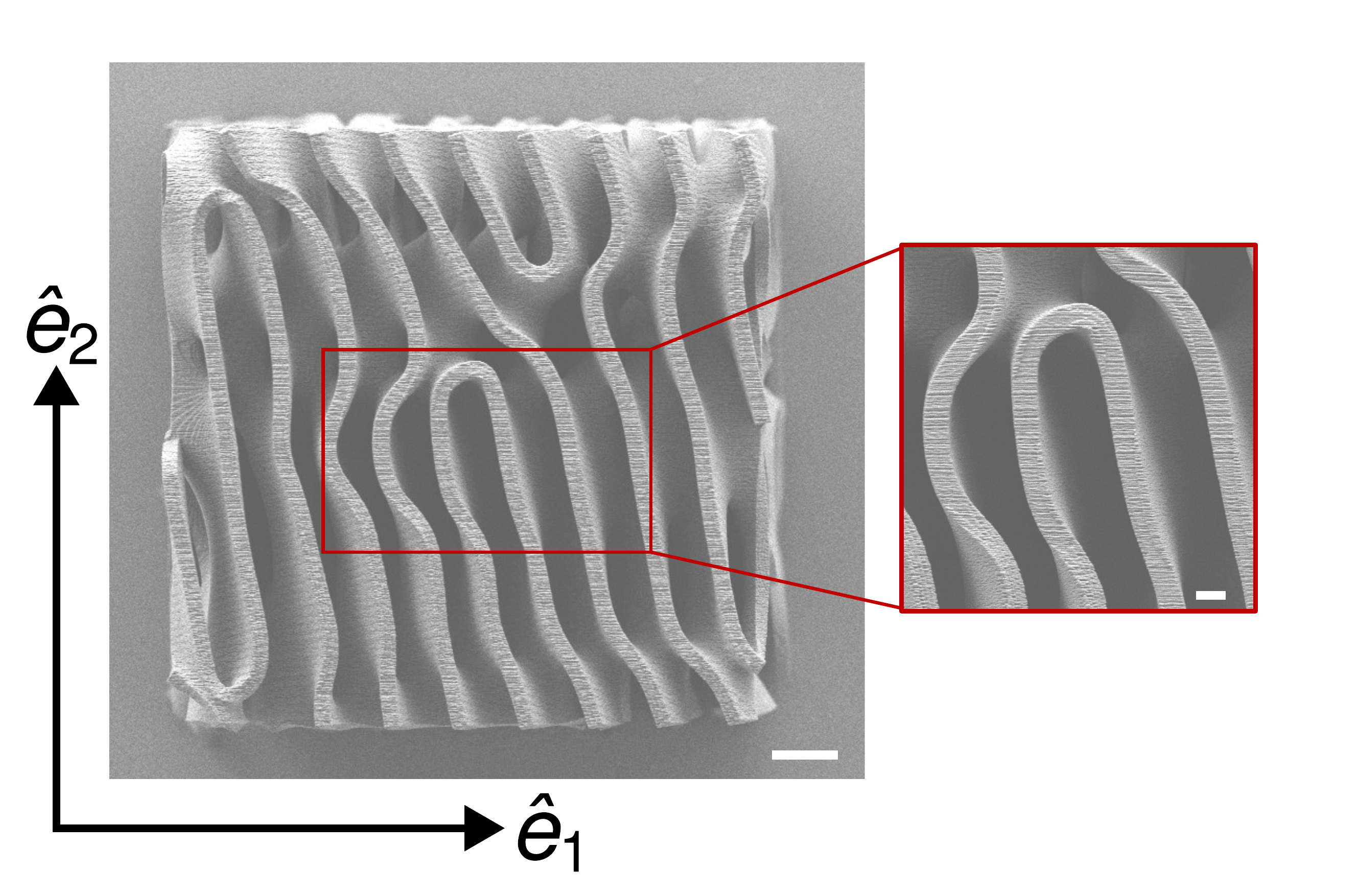}
	\centering
	\caption{SEM image of sample $\bfTheta = [0^\circ,0^\circ,33^\circ]$ which shows the print quality of the thinnest regions of the samples. Scale bars, 10 \textmu{}m  and 2 \textmu{}m (inset).}
	\label{fig:Defeacts}
\end{figure}

We  performed uniaxial quasi-static \emph{in situ} and \emph{ex situ} compression experiments using a 400-\textmu{}m flat punch tip (Synton-MDP AG) and a displacement-controlled nanoindenter (Alemnis AG). All experiments consisted of a compression to 50 \textmu{}m. A strain rate of $10^{-3}$ s$^{-1}$ was maintained for all experiments. The \emph{in situ} compressions were performed in a Zeiss SEM (Sigma HD VP) under the same testing conditions as the \emph{ex situ} experiments except to a compression displacement of 40 \textmu{}m . The stress-strain data was calculated by normalizing the load data with the mean cross-sectional area and height of each sample batch.
We present the \emph{ex situ} dataset past 40\% strain with the unloading region in \figurename~\ref{fig:all_experiments}, along with detailed \emph{in situ} responses in Figures~\ref{fig:Laminsitu}--\ref{fig:cubicnsitu}. Further, \figurename~\ref{fig:Defeacts} demonstrates the print quality of all samples, where a $\bfTheta = [0^\circ,0^\circ,33^\circ]$ sample is used to illustrate the thinnest sections of the spinodal architectures. 

\begin{figure}[ht]
	\includegraphics[width = 1.0\textwidth]{Figure_S6.pdf}
	\centering
	\caption{\emph{In situ} compression experiments for $\bfTheta=[0^\circ,0^\circ,33^\circ]$. ({\textbf{a}}) Normalized stress-strain response for the loading direction $\hat{\bfe}_1$ (pink), $\hat{\bfe}_2$ (blue), and $\hat{\bfe}_3$ (orange). The $\hat{\bfe}_1$ loaded curve has the strain points of 0\% (i), 20\% (ii), and 40\% (ii) indicated to match the compression snapshots. (\textbf{b}), (\textbf{c}), and (\textbf{d}) Snapshots of the compression experiment for the loading direction of $\hat{\bfe}_1$, $\hat{\bfe}_2$ , and $\hat{\bfe}_3$, respectively. Scale bars, 20 \textmu m. } 
	\label{fig:Laminsitu}
\end{figure}

\begin{figure}[ht]
	\includegraphics[width = 1.0\textwidth]{Figure_S7.pdf}
	\centering
	\caption{\emph{In situ} compression experiments for $\bfTheta=[0^\circ,23^\circ,37^\circ]$.({\textbf{a}}) Normalized stress-strain response for the loading direction $\hat{\bfe}_1$ (pink), $\hat{\bfe}_2$ (blue), and $\hat{\bfe}_3$ (orange). The $\hat{\bfe}_1$ loaded curve has the strain points of 0\% (i), 20\% (ii), and 40\% (ii) indicated to match the compression snapshots. (\textbf{b}), (\textbf{c}), and (\textbf{d}) Snapshots of the compression experiment for the loading direction of $\hat{\bfe}_1$, $\hat{\bfe}_2$ , and $\hat{\bfe}_3$ respectively. Scale bars, 20 \textmu m. } 
	\label{fig:Colinsitu}
\end{figure}

\begin{figure}[ht]
	\includegraphics[width = 1.0\textwidth]{Figure_S8.pdf}
	\centering
	\caption{\emph{In situ} compression experiments $\bfTheta=[20^\circ,23^\circ,28^\circ]$.({\textbf{a}}) Normalized stress-strain response for the loading direction $\hat{\bfe}_1$ (pink), $\hat{\bfe}_2$ (blue), and $\hat{\bfe}_3$ (orange). The $\hat{\bfe}_1$ loaded curve has the strain points of 0\% (i), 20\% (ii), and 40\% (ii) indicated to match the compression snapshots. (\textbf{b}), (\textbf{c}), and (\textbf{d}), Snapshots of the compression experiment for the loading direction of $\hat{\bfe}_1$, $\hat{\bfe}_2$ , and $\hat{\bfe}_3$ respectively. Scale bars, 20 \textmu m. } 
	\label{fig:cubicnsitu}
\end{figure}

\subsection*{Computing the absorbed energy and incremental stiffness}
Given a stress-strain response $\{(\varepsilon_t,\sigma_t):t=1,\dots,T\}$, we define the absorbed energy at loadstep $t$ as the cumulative area under the stress-strain curve and compute it using the trapezoidal rule as
\be
\text{Absorbed energy at loadstep $t$} =  \sum_{\tau=1}^{t}\frac{(\sigma_\tau+\sigma_{\tau-1})}{2}(\varepsilon_\tau-\varepsilon_{\tau-1}),\qquad \text{with}\quad \sigma_0=\varepsilon_0 =0.
\ee
We define the incremental stiffness as the slope of the stress-strain curve. Assuming uniform spacing of strain points equal, we compute the incremental stiffness using the second-order central differences (and first order finite difference when necessary) as
\be
\text{Incremental stiffness at loadstep $t$} = 
\begin{cases}
    {(\sigma_{t+1} - \sigma_{t-1}})/{(\varepsilon_{t+1} - \varepsilon_{t-1})} \quad &\text{if} \quad t>1\  \text{and} \ t<T\,\\
    {(\sigma_1 - \sigma_0)}/{(\varepsilon_1 - \varepsilon_0)} \quad &\text{if} \quad t=1,\\
    {(\sigma_T - \sigma_{T-1})}/{(\varepsilon_{T} - \varepsilon_{T-1})} \quad &\text{if} \quad t=T.
\end{cases}
\ee
\subsection*{Finite Element Analysis}

\begin{figure}[h!]
    \centering
    \includegraphics[width = 0.75\textwidth]{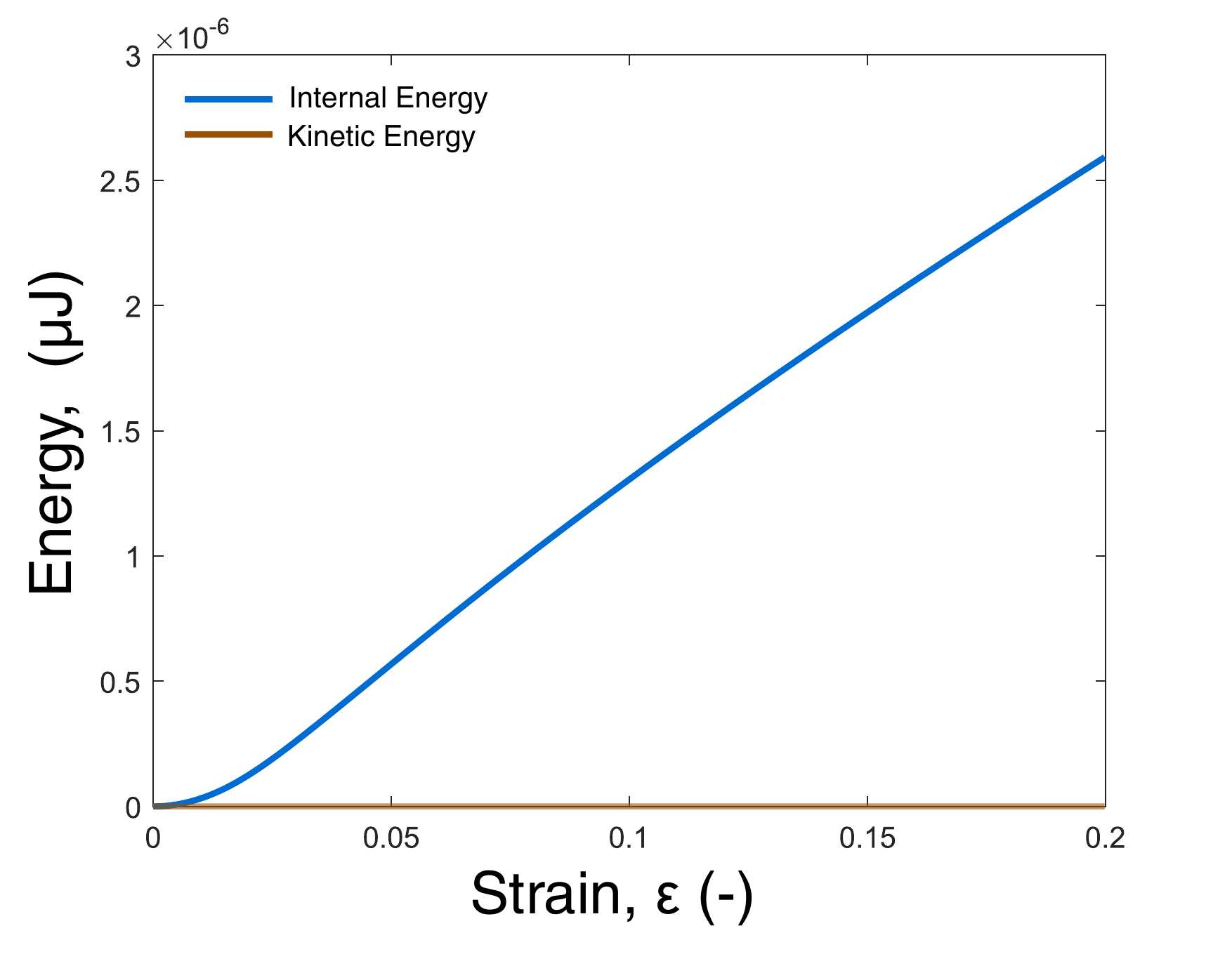}
    \caption{Magnitude of the kinetic energy compared to the internal energy of our finite element simulations in (\textmu{}J). The graph shows that the kinetic energy was negligible over the entire strain range of our simulation, verifying the quasi-static loading boundary condition.}
    \label{fig:KE_vs_IE}
\end{figure}

The finite element analysis was performed in Abaqus/Explicit. The simulations accounted for contact, geometric nonlinearities, and employed a nonlinear material model. The self-contact was modeled by creating a general contact definition within Abaqus interaction settings. A frictional penalty was assigned to the whole body with a coefficient of friction of 0.4 for the tangential friction interaction behavior while hard contact was used for the normal direction. Displacement boundary conditions along the compression direction were set at the nodes on the top face with the other displacements constrained and rotations unset. 
All degrees of freedom of the nodes on the bottom face were fully constrained. The degrees of freedom of lateral faces were left unconstrained, matching the conditions in experiments. To ensure quasi-static conditions were maintained, we ensured the kinetic energy of the system was a negligible fraction of the internal energy (\figurename~\ref{fig:KE_vs_IE}). A strain rate of $\dot{\varepsilon}=10^{-3}$ s\textsuperscript{-1} was imposed, and a fixed mass scaling factor of 100 was applied. All other simulation parameters are listed in \tablename~\ref{tab:sim_parameters}.

\begin{figure}[h!]
    \centering
    \includegraphics[width = 0.75\textwidth]{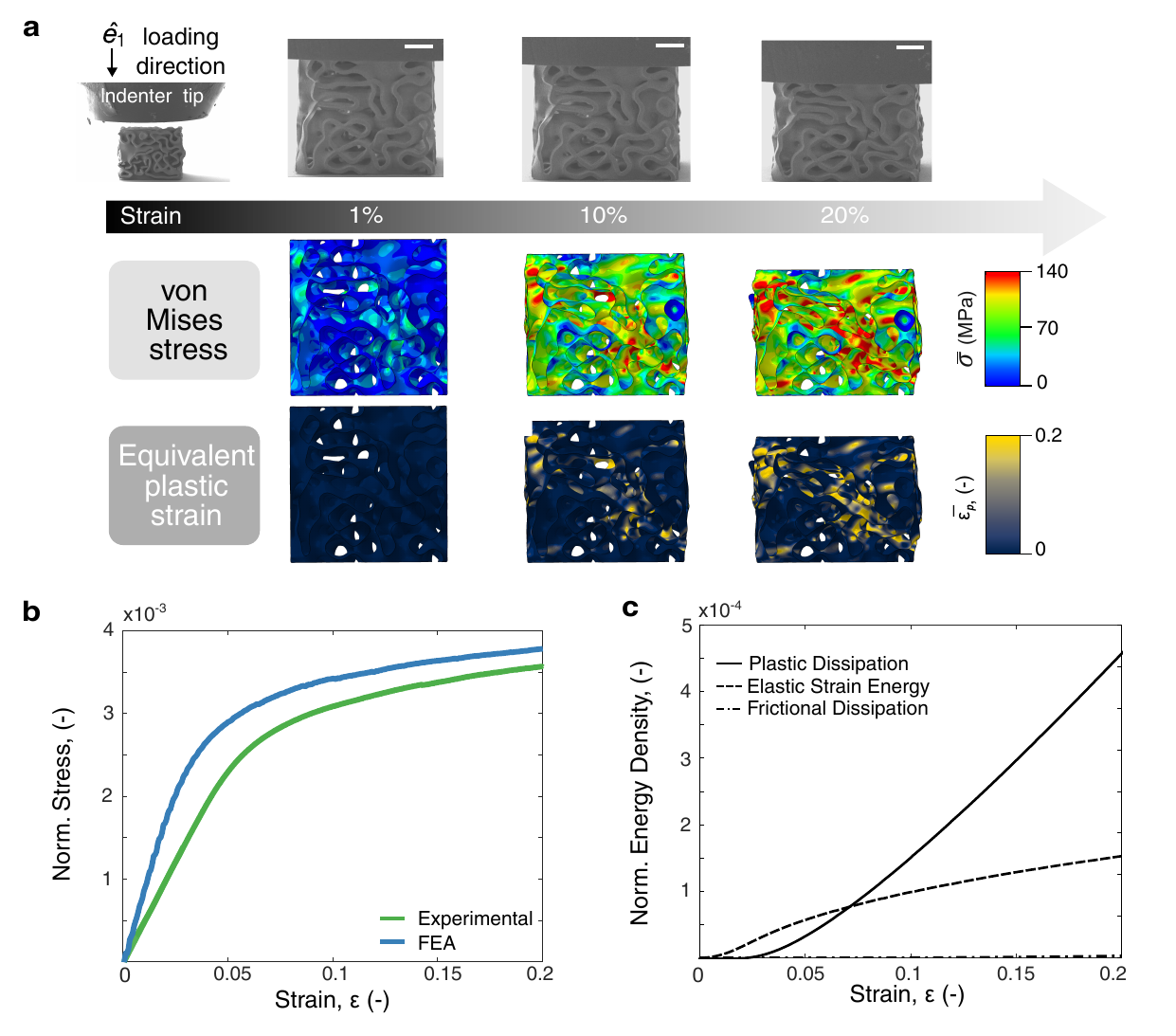}
    \caption{Mesh convergence results for $\bfTheta=[0^\circ,23^\circ,37^\circ]$ loaded int he $\hat{\bfe}_1$ direction. Plasticity behavior converges around 200k S3 elements.  }
    \label{fig:Mesh_Conv}
\end{figure}

A mesh convergence study was then performed on the $\bfTheta=\{0,23^\circ,37^\circ\}$ architecture. The meshes were edited in MeshLab were a simplification filter was applied followed by multiple applications of the Taubin smoothing filter. The resulting mesh was imported into Abaqus and meshed with S3 elements. For the study, the sample was compressed along $\hat{\bfe}_1$, reaching convergence at $\sim$ 200,000 elements (\figurename~\ref{fig:Mesh_Conv}).

\begin{table}[h!]
\centering
\caption{Parameters for spinodal compression simulations. } 
\label{tab:sim_parameters}
\begin{tabular}{lcc}
\hline
\multicolumn{1}{l}{\textbf{Parameter}} & \textbf{Notation} & \textbf{Value} \\ \hline
Young's modulus &$E$ &  3200 MPa \\
Yield Strength &$\sigma_y$ & 77 MPa \\
Density & $\rho$ & 1200 kg/m\textsuperscript{3} \\
Poisson's ratio & $\nu$ & 0.36 \\
Shell thickness & $t$ & 3 \textmu{}m \\
Friction coefficient & $f_c$ & 0.4 \\
Mass scaling factor & $M$ & 100
 \\\hline
\end{tabular}
\end{table}

\begin{figure}[h!]
    \centering
    \includegraphics[width = 0.75\textwidth]{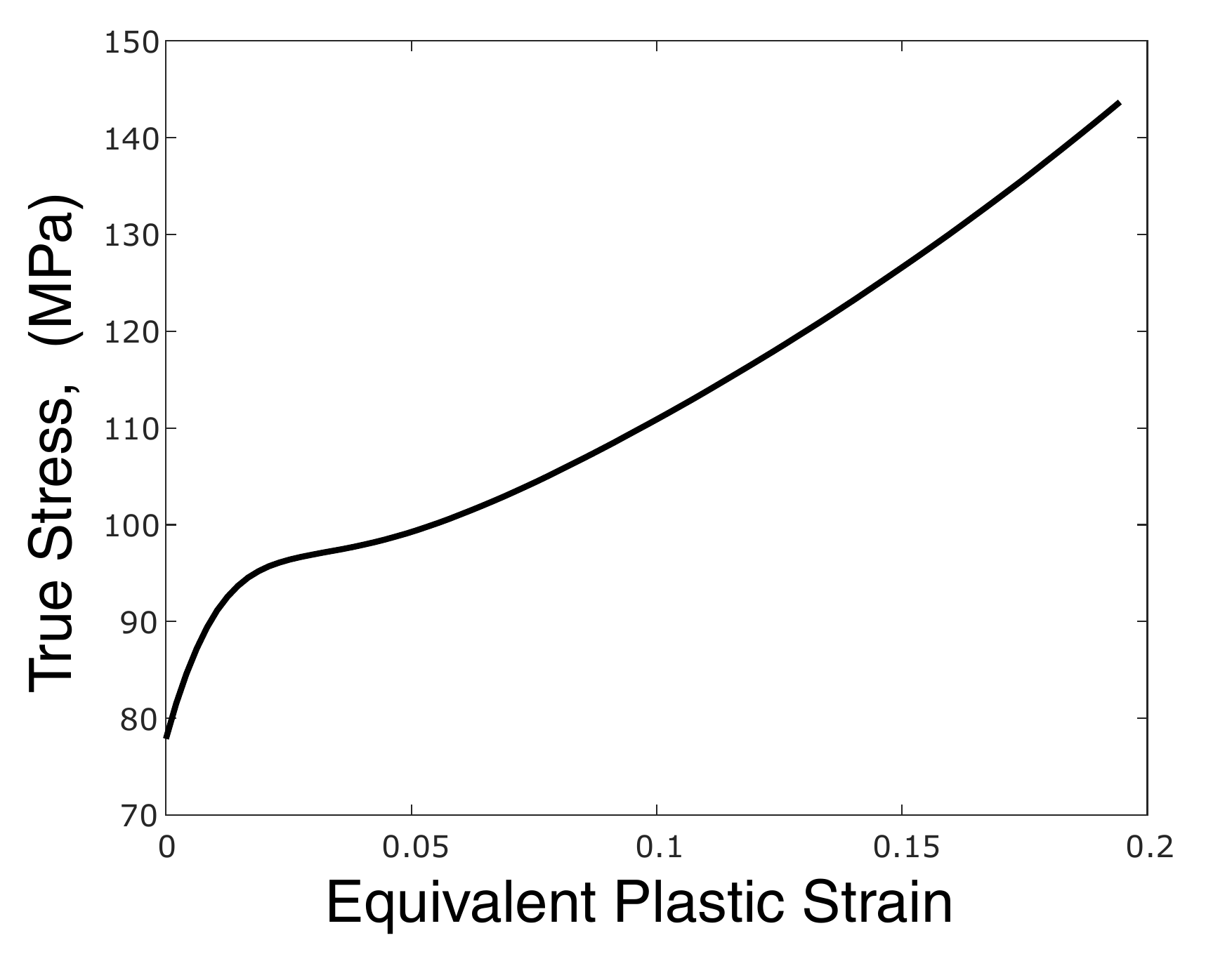}
    \caption{The plastic flow curve used to input the isotropic hardening plasticity parameters in Abaqus for our elastic-plastic material model.}
    \label{fig:Flow_curve}
\end{figure}

To model plasticity in our elasto-plastic material model, we used the experimental flow hardening curve data from averaged engineering stress-strain curves of 20 randomly selected monolithic IP-Dip pillars under compression, which was input into Abaqus as isotropic hardening plastic material parameters. This flow hardening curve was obtained by first taking the averaged engineering stress-strain response of IP-Dip and converting it to true stress-strain, assuming incompressibility of plastic deformation. While this assumption has its basis on plastic deformation of crystalline materials such as metals, we employ it here as a simplistic approximation of plastic deformation in our polymers. 
This results in a true stress definition of
\begin{equation}
    \sigma_T=\sigma(1-e),
\end{equation}
where $\sigma$ is the engineering stress and $e$ is the engineering strain. Taking the data beyond the yield strength of the material (found from the 0.2\% offset, 77 MPa) yielded the flow stress vs. equivalent plastic strain relation. \figurename~\ref{fig:Flow_curve} shows the resulting isotropic hardening flow curve parameters used for our finite element simulations. 
\newpage

\begin{figure}[h!]
    \centering
    \includegraphics[width = 0.75\textwidth]{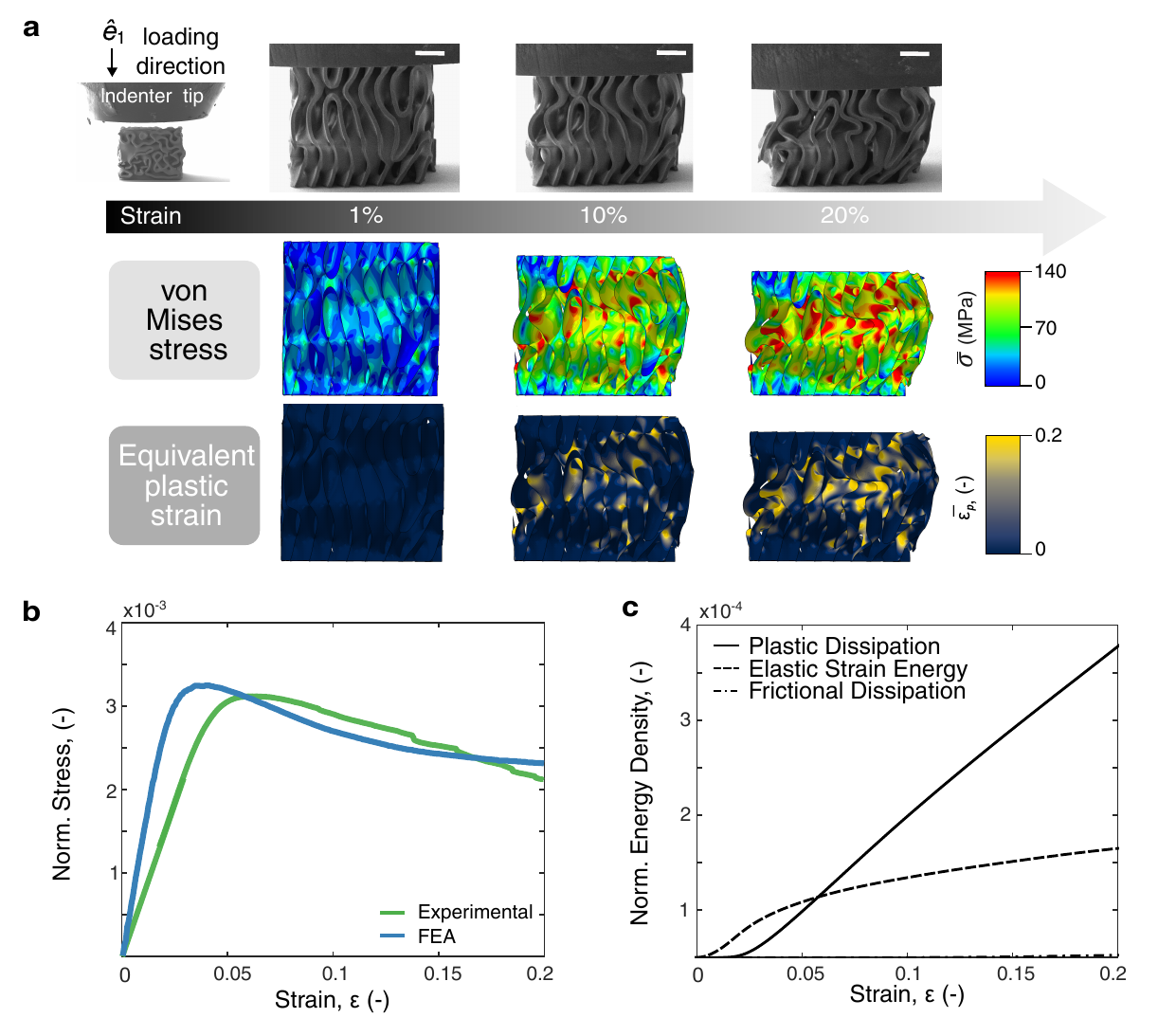}
    \caption{(\textbf{a}) \emph{In situ} snapshots for $\bfTheta=[0^\circ,0^\circ,33^\circ]$ loaded in the $\hat{\bfe}_1$ direction for $1\%$, $10\%$, and $20\%$ strain points. Scale bars, 10 \textmu{}m. (\textbf{b}) Comparison between \emph{in situ} experiment (green line) and simulation (blue line) for the normalized stress-strain curves up to 20\% strain. (\textbf{c}) Distribution of energy mechanisms as a function of strain, including  plastic dissipation, elastic strain energy, and frictional dissipation. Up to 20 \% strain, friction effects are negligible and plastic dissipation is the dominant mode of dissipation. }
    \label{fig:SI_FEM_33}
\end{figure}
\begin{figure}[h!]
    \centering
    \includegraphics[width = 0.75\textwidth]{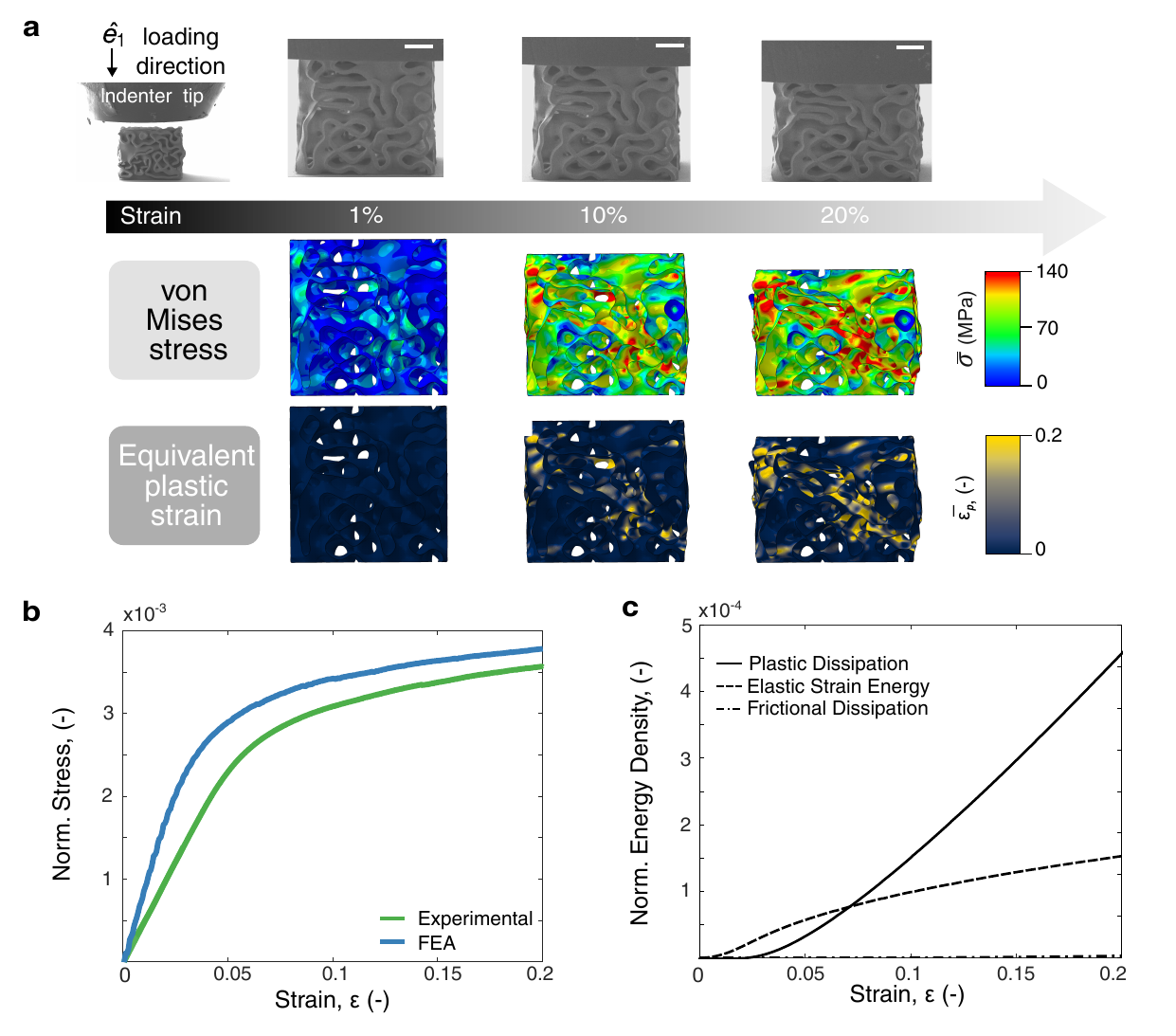}
    \caption{(\textbf{a}) \emph{In situ} snapshots for $\bfTheta=[20^\circ,23^\circ,28^\circ]$ loaded in the $\hat{\bfe}_1$ direction for $1\%$, $10\%$, and $20\%$ strain points. Scale bars, 10 \textmu{}m. (\textbf{b}) Comparison between \emph{in situ} experiment (green line) and simulation (blue line) for the normalized stress-strain curves up to 20\% strain. (\textbf{c}) Distribution of energy mechanisms as a function of strain, including  plastic dissipation, elastic strain energy, and frictional dissipation. Up to 20 \% strain, friction effects are negligible and plastic dissipation is the dominant mode of dissipation.}
    \label{fig:SI_FEM_202328}
\end{figure}
\newpage
\section*{Supplementary Note 3. Machine learning framework }
\subsection*{Data preprocessing}
All the non-zero cones angles are sampled within the range of $[\theta_{\min},\theta_{\max}$] and are subsequently mapped within the range of $[0,1]$ by dividing all the angles by $\theta_{\max}$. Moreover, the stress-strain data obtained from the micromechanical experiments are linearly interpolated to $T$ fixed and uniformly-spaced strain values between $\varepsilon_1=0.001$ and $\varepsilon_T=0.4$, i.e., $\{\varepsilon_1, \varepsilon_1 + \Delta\varepsilon,\varepsilon_1+2\Delta\varepsilon,\dots,\varepsilon_T\}$ (with constant $\Delta\varepsilon$; see \tablename \ \ref{tab:datagen_parameters} for value used) for consistent training and testing data. Lastly, all the stress-strain data pairs are normalized to the range of $[0,1]$ by dividing the stress and strains values by their respective maximums encountered in the training dataset (across all directions).
We normalize all quantities (i.e., stress, absorbed energy, and incremental stiffness), by dividing them by the elastic modulus $E_s$ of the constituent IP-Dip photoresist.

\subsection*{Derivation of stress}
Here, we show the derivation of stress in equation (5) of the main article. Recall the nonconvex energy potential formulation from equation (4) of the main article.
\be\label{eq:Wrelax-supp}
W(\varepsilon,\bfTheta) = \min_{\gamma_1,\gamma_2}\  \Bigg[\sum_{i=1}^2 \gamma_i W_i(\varepsilon,\bfTheta)  - \underbrace{\left( - k_T \sum_{i=1}^2\gamma_i \log \gamma_i \right)}_{\text{configurational entropy}}\Bigg], \qquad \text{with} \quad \gamma_1 + \gamma_2 = 1.
\ee
To find the minimizers $\gamma_1^*$ and $\gamma_2^*$ in \eqref{eq:Wrelax-supp}, we first substitute
\be
\gamma_2 = 1 - \gamma_1,
\ee
such that the minimization problem simplifies as follows:
\be
W(\varepsilon,\bfTheta) = \min_{\gamma_1}\  \Bigg[\gamma_1 W_1(\varepsilon,\bfTheta) + (1-\gamma_1) W_2(\varepsilon,\bfTheta)  - \left( - k_T \left[ \gamma_1 \log \gamma_1 + (1-\gamma_1)\log(1-\gamma_1)\right] \right)\Bigg].
\ee
We solve for the minimizers by setting the derivative to 0, i.e.,
\be
\partderiv{W(\varepsilon,\bfTheta)}{\gamma_1} = W_1(\varepsilon,\bfTheta) -W_2(\varepsilon,\bfTheta)  - \left( - k_T \left[ \log \gamma_1^* + 1 - \log(1-\gamma_1^*) - 1\right] \right) = 0.
\ee
Solving the above for $\gamma_1^*$ yields
\be
\gamma_1^*=\frac{1}{\exp\left(-(W_2 - W_1)/k_T\right) + 1}.
\ee
Substituting $\gamma_1^*$ and $\gamma_2^*=(1-\gamma_1^*)$  in \eqref{eq:Wrelax-supp} and computing ${\partial W(\varepsilon,\bfTheta)}/{\partial \varepsilon}$ gives the analytical form of $\sigma(\varepsilon,\bfTheta)$ in equation (5) of the main article.
\clearpage
\subsection*{PICNN architecture details}
The PICNN framework was introduced by Amos et al \cite{amos2017input}. and is an extension to the Input Convex Neural Network (ICNN). The main difference is that the output of the PICNN is only convex with respect to some inputs, but has arbitrary functional relationships with respect to the remaining inputs. Here, we describe the layer-wise architecture of the PICNN in the scope of this work; refer to \tablename \ \ref{tab:parameters} for the list of hyperparameter values. 
\begin{subequations}
    \small
    \begin{align}
        \text{Nonconvex Input}: & \quad  \bfu_0 = \bfTheta^{(n)}\\
        \text{Convex Input}: & \quad  \bfz_0=\varepsilon_i^{(t,n)}\\
        \text{First nonconvex hidden layer of size $d_1$ }: & \quad \bfu_1 = \Tilde{\calG}\left(\bfA_{uu}^{(0)}\bfu_0+\bfb^{(0)}_{uu}\right)\\
        &\quad \text{with trainable parameters:}\ 
        \bfA_{uu}^{(0)}\in\Rset^{d_1\times\left|\bfu_{0}\right|},\  
        \bfb_{uu}^{(0)}\in\Rset^{d_1}\notag\\
        \text{First convex hidden layer of size $d_1$ }: & \quad \bfz_1 = \calG\left(\hat\calG({\bfA_{zz}^{(0)}})\left(\bfz_0\odot\left[\bfA_{zu}^{(0)}\bfu_0+\bfb_{zu}^{(0)}\right]_+\right)+\bfA_{u}^{(0)}\bfu_0+\bfb_{u}^{(0)}\right)\\
        &\quad \text{with trainable parameters:}\notag \\
        &\quad \bfA_{zz}^{(0)}\in\Rset^{d_1\times\left|\bfz_0\right|},
        \  
        \bfA_{zu}^{(0)}\in\Rset^{\left|\bfz_0\right|\times\left|\bfu_0\right|},
        \bfb_{zu}^{(0)}\in\Rset^{\left|\bfz_0\right|}, \notag\\
        &\quad \bfA_{u}^{(0)}\in\Rset^{d_1\times\left|\bfu_0\right|},
        \bfb_{u}^{(0)}\in\Rset^{d_1}, \notag\\
        \vdots\qquad \qquad & \notag\\
        \text{$(k+1)^\text{th}$ nonconvex hidden layer of size $(d_{k+1})$}: & \quad \bfu_{k+1} = \Tilde{\calG}\left(\bfA_{uu}^{(k)}\bfu_k+\bfb^{(1)}_{uu}\bfu_k\right)\\
        &\quad \text{with trainable parameters:}\ 
        \bfA_{uu}^{(k)}\in\Rset^{d_{k+1}\times d_{k}},\  
        \bfb_{uu}^{(k)}\in\Rset^{d_{k+1}}\notag\\
        \text{$(k+1)^\text{th}$ convex hidden layer of size $(d_{k+1})$ }: & \quad \bfz_{k+1} = \calG\bigg(\hat\calG(\bfA_{zz}^{(k)})\left(\bfz_k\odot\left[\bfA_{zu}^{(k)}\bfu_k+\bfb_{zu}^{(k)}\right]_+\right)\label{eq:(k+1)-hidden-z}  \\
        & \quad\quad\quad\quad +\underbrace{\bfA_{yy}^{(k)}\left(\bfz_0\odot\left(\bfA_{yu}^{(k)}\bfu_k+\bfb^{(k)}_{yu}\right)\right)+b_{yy}^{(k)}}_{\text{skip-connection}}+\bfA_{u}^{(k)}\bfu_k+\bfb_{u}^{(k)}\bigg)\notag\\
        &\quad \text{with trainable parameters:}\notag \\
        &\quad \bfA_{zz}^{(k)}\in\Rset^{d_{k+1}\times d_{k}},
        \bfA_{zu}^{(k)}\in\Rset^{d_k\times d_k},
        \bfb_{zu}^{(k)}\in\Rset^{d_k}, \notag\\
        &\quad \bfA_{u}^{(k)}\in\Rset^{d_{k+1}\times d_k},
        \bfb_{u}^{(k)}\in\Rset^{d_{k+1}},\bfA_{yy}^{(k)}\in\Rset^{d_{k+1}\times \left|\bfz_0\right|},
        \bfb_{yy}^{(k)}\in\Rset^{d_{k+1}},
        \notag\\
        &\quad \bfA_{yu}^{(k)}\in\Rset^{\left|\bfz_0\right|\times d_k},
        \bfb_{yu}^{(k)}\in\Rset^{\left|\bfz_0\right|}, \notag\\
        \vdots\qquad \qquad & \notag\label{eq:last-hidden-z}\\
        \text{Output ($L^\text{th}$) layer of size $d_L=1$}: & \quad  \text{Same as \eqref{eq:(k+1)-hidden-z} with $k=(L-1)$, \ $\bfA^{(L-1)}_u = \bm{0}$, and \ $\bfb^{(L-1)}_u = \bm{0}$. }
    \end{align}
\end{subequations} 
Here, $\Tilde\calG$, $\hat{\calG}$ and $\calG$ denote the following activation functions (commonly known as, rectified linear unit (ReLU), softplus, and squared softplus, respectively)
\be
\Tilde\calG = \max(x,0), \quad \hat\calG = \log(1+\exp(x)), \quad \calG = \frac{1}{\alpha}\left(\log(1+\exp(x)\right)^2.
\ee
The brackets $[\cdot]_+$ also refer to the $\tilde\calG$ activation.

\subsection*{Training protocols}
The optimized dimensions and hyperparameters (including the number of hidden layers, nodes in each layer, activation functions, learning rates, etc.) for the PICNN-based framework and the property predictor are detailed in the Supplementary Table \ref{tab:parameters}. The PICNN-based framework is divided into three separate submodules for each direction $\hat{\bfe}_1,\hat{\bfe}_2$, and $\hat{\bfe}_3$. Furthermore, each submodule consists of three pairs of PICNNs for each case of spinodoid morphology (lamellar, columnar, and cubic) and one pair of MLPs for predicting the $(b,v)$-offsets. The input dimensions vary depending on the corresponding spinodal morphology (see \tablename~\ref{tab:parameters}). Each model within a submodule is trained jointly. We train each submodule for $3,000$ epochs with a learning rate of $0.005$ and the Adam optimizer\cite{kingma2014adam}. We modify the loss function (Equation (9) in the main article) to return $0$ if there is a division by $0$ (i.e., the true stress value is $0$). The training is performed on an NVIDIA RTX A5000. Moreover, we use a transfer learning technique to improve the accuracy of the models for the $\hat{\bfe}_1$-direction, namely: (i) we first train the model in the $\hat{\bfe}_2$-direction, (ii) we use the model weights in $\hat{\bfe}_2$-direction as a starting point to train the model in $\hat{\bfe}_1$-direction. Since the responses in $\hat{\bfe}_1$- and $\hat{\bfe}_2$-direction are bound to be similar in certain cases (due to the ascending ordering of the angles), the pre-training helps improve the learning process of the model.

\begin{table}[ht]
\centering
\caption{List of parameters used for the ML training and design optimization protocols.} 
\label{tab:parameters}
\begin{tabular}{lcc}
\hline
\multicolumn{1}{l}{\textbf{Parameter}} & \textbf{Notation} & \textbf{Value} \\ \hline
\textit{ML framework hyperparameters:}\\
$\quad$ Input dimension convex path PICNN &$-$ &  $1$ \\
$\quad$ Input dimension nonconvex path PICNN lamellar &$-$ &  $1$ \\
$\quad$ Input dimension nonconvex path PICNN columnar &$-$ &  $2$ \\ 
$\quad$ Input dimension nonconvex path PICNN cubic & $-$ & $3$ \\
$\quad$ Number of hidden layers PICNN  & $L$ & $3$ \\
$\quad$ Hidden dimensions lamellar & $d_1,d_2,d_3$ & $16,16,16$ \\
$\quad$ Hidden dimensions columnar & $d_1,d_2,d_3$ & $32,32,32$ \\
$\quad$ Hidden dimensions cubic & $d_1,d_2,d_3$ & $48,48,48$ \\
$\quad$ Output dimension & $-$ &$1$ \\
$\quad$ Learning rate &$-$ & $0.005$ \\
$\quad$ Squared softplus scaling &$\alpha$ & $\frac{1}{d_L}$ \\
$\quad$ Batch size &$-$ & $512$ \\
$\quad$ Optimizer &$-$ & Adam\cite{kingma2014adam}\\\hline
\textit{Optimization setup hyperparameters:}\\
$\quad$ Number of guesses & $S$    & $10$\\
$\quad$ Number of epochs & $E$    & $30$  \\
$\quad$ Target curve multiplier & $\kappa$    & $1.2$  \\
$\quad$ Optimizer &$-$ & Adam\cite{kingma2014adam}\\
\end{tabular}
\end{table}

\section*{Supplementary Note 4. Optimization framework details}
All hyperparameters regarding the optimization can be found in \tablename \ \ref{tab:parameters}. For inverse design, we choose a random target curve from the dataset and multiply it by $\kappa$ to obtain a target curve which is outside the training regime. Since we design the stress-strain response for a single direction under uniaxial compression, we use all nine possible cases (see \tablename \ \ref{tab:NN_cases}) to generate the optimal design parameters to match the queried stress-strain curve. Hence, for each case, we perform a multi-initialization optimization, meaning, we perform $9S$ total gradient-based optimizations to find the optimal design. In Algorithm~\ref{alg:optimization} we provide a pseudocode of the optimization scheme. For each of the $9S$ initial guesses, we take a uniformly distributed sample in the range $[\theta_{\min},\theta_{\max}]$ for the design parameters $\bfTheta$. Depending on the case of the spinodal morphology type, we either sample 1,2, or 3 angles. In each epoch, the  design parameters in $\bfTheta$ are first ordered in the ascending order of $\theta_1\leq\theta_2\leq\theta_3$  before passing them to the forward ML models as inputs (recall the permutation invariance of the design parameterization for spinodal metamaterials; see Supplementary Information Note 1). The mean absolute percentage error is computed between the entire predicted and target stress-strain curve. We employed the Adam optimizer to update the design parameters. After $E$ epochs, we compare the final loss to the current best loss. If the loss of the current guess performed better, we save them as the best set of design parameters $\bfTheta^*$, otherwise, we continue with the next guess. Though this process is shown in serial, the optimization can also be done in parallel.

\begin{table}[ht]
\caption{For each loading direction $\{\hat{\bfe}_1,\hat{\bfe}_2,\hat{\bfe}_3\}$ we have $8$ jointly trained models, each with separate weights and corresponding to the three morphologies, namely, lamellar, columnar, and cubic. Combined with all loading directions, this results in 18 PICNNs and 6 MLPS.}
\centering
\begin{tabular}{@{}ccccc@{}}
\toprule
\textbf{\#} & \textbf{model component} & \textbf{Architecture} & \textbf{Input} & \textbf{Output} \\ \midrule
1       & lamellar (1)  & PICNN &$\tilde\varepsilon_t^{(n,i)},\bfTheta^{(n)}$ &$W_1$               \\
2       & lamellar (2)  & PICNN &$\left(\tilde\varepsilon_t^{(n,i)}-b\right),\bfTheta^{(n)}$ &$W_2$            \\
3       & columnar (1)  & PICNN &$\tilde\varepsilon_t^{(n,i)},\bfTheta^{(n)}$ &$W_1$               \\
4       & columnar (2)  & PICNN & $\left(\tilde\varepsilon_t^{(n,i)}-b\right),\bfTheta^{(n)}$&$W_2$              \\
5       & cubic (1)  & PICNN &$\tilde\varepsilon_t^{(n,i)},\bfTheta^{(n)}$ &$W_1$              \\
6       & cubic (2)  & PICNN & $\left(\tilde\varepsilon_t^{(n,i)}-b\right),\bfTheta^{(n)}$&$W_2$               \\
7       & $b$-net  & MLP & $\bfTheta^{(n)}$ & $b$                \\
8       & $v$-net  & MLP & $\bfTheta^{(n)}$ & $v$              \\
\bottomrule

\end{tabular}

\label{tab:NN_cases}
\end{table}

\begin{algorithm}[t]
\caption{Gradient-based optimization for target stress-strain response}\label{alg:optimization}
\begin{algorithmic}[1]
\State \textbf{Input}: Target stress-strain response $\{(\hat\varepsilon_t,\hat\sigma_t):t=1,\dots,T\}$
\State $\text{best\_loss}\gets +\infty$\Comment{Best loss}
\State $\bfTheta^*\gets [0,0,0]$\Comment{Best design parameters}
\State $\hat\bfe^*\gets [0,0,0]$\Comment{Best loading direction}
\For{$s$ in $1,\dots,S$} \Comment{Iterate over different guesses}
\For{$\hat{\bfe}_i$ in $\{\hat{\bfe}_1,\hat{\bfe}_2,\hat{\bfe}_3$\}} \Comment{Iterate over different loading directions}
\For{$c$ in \{lamellar, columnar, cubic\}} \Comment{Iterate over different spinodal morphology types}
\If{$c$ = lamellar}   \Comment{Uniformly sample 1,2 or 3 angles depending on spinodal morphology type}
    \State $\bfTheta \sim \calU([\theta_\text{min},\theta_\text{max}])$
\ElsIf{$c$ = columnar} 
    \State $\bfTheta \sim \calU([\theta_\text{min},\theta_\text{max}]\times[\theta_\text{min},\theta_\text{max}])$
\Else
    \State $\bfTheta \sim \calU([\theta_\text{min},\theta_\text{max}]\times[\theta_\text{min},\theta_\text{max}]\times[\theta_\text{min},\theta_\text{max}])$
\EndIf

\State $\text{loss} \gets 0$ \Comment{Initialize loss variable}
\For{$e$ in $1,\dots,E$} \Comment{Iterate over epochs}
\State $\text{loss} \gets 0$ \Comment{Reset loss variable}
\State ${\bfTheta} \gets $ sort$({\bfTheta})$ \Comment{Sort the angles in ascending order}

\For{$t$ in $1,\dots,T$} \Comment{Iterate over loadsteps}
\State $\Tilde{\sigma}_t \gets \sigma^{(c,i)}\left(\hat\varepsilon_t,{\bfTheta}\right)$\Comment{Use ML model associated with direction $\hat{\bfe}_i$ and case $c$}
\State $\text{loss} \gets \text{loss} + \frac{1}{T}\left|\frac{\Tilde{\sigma}_t-\hat\sigma_t}{\hat\sigma_t}\right|$ \Comment{Compute loss}
\EndFor
\State $\bfTheta \gets \text{Adam}\left(\bfTheta,\text{loss}\right)$ \Comment{Update $\bfTheta$ with Adam\cite{kingma2014adam} optimizer}
\EndFor
\If{$\text{loss} < \text{best\_loss}$} \Comment{Update best guess if smaller loss was achieved}
    \State $\text{best\_loss} \gets \text{loss}$
    \If{$c$ = lamellar}   \Comment{Update $\bfTheta^*$ depending on spinodal morphology type}
    \State $[\Theta^*_0,\Theta^*_1] \gets [0,0]$
    \State $[\Theta^*_2] \gets \bfTheta$
    \ElsIf{$c$ = columnar} 
    \State $[\Theta^*_0] \gets [0]$
    \State $[\Theta^*_1,\Theta^*_2] \gets \bfTheta$
    \Else \State $[\Theta^*_1,\Theta^*_2,\Theta^*_3] \gets \bfTheta$
    \EndIf
    \State $\hat\bfe^*\gets\hat\bfe_i$ \Comment{Update best loading direction}
\EndIf
\EndFor
\EndFor
\EndFor
\State \Return  $\bfTheta^*$, $\hat\bfe^*$
\end{algorithmic}
\end{algorithm}

\clearpage

\section*{Movie S1.}
\textit{In situ} compression of the lamellar morphology, $\bfTheta=[0^\circ,0^\circ,33^\circ]$, loaded in the $\hat{\bfe}_1$ direction with the corresponding stress-strain curve. 

\section*{Movie S2.}
\textit{In situ} compression of the lamellar morphology, $\bfTheta=[0^\circ,0^\circ,33^\circ]$, loaded in the $\hat{\bfe}_2$ direction with the corresponding stress-strain curve. 

\section*{Movie S3.}
\textit{In situ} compression of the lamellar morphology, $\bfTheta=[0^\circ,0^\circ,33^\circ]$, loaded in the $\hat{\bfe}_3$ direction with the corresponding stress-strain curve. 

\section*{Movie S4.}
\textit{In situ} compression of the columnar morphology, $\bfTheta=[0^\circ,23^\circ,37^\circ]$, loaded in the $\hat{\bfe}_1$ direction with the corresponding stress-strain curve. 

\section*{Movie S5.}
\textit{In situ} compression of the columnar morphology, $\bfTheta=[0^\circ,23^\circ,37^\circ]$, loaded in the $\hat{\bfe}_2$ direction with the corresponding stress-strain curve. 

\section*{Movie S6.}
\textit{In situ} compression of the columnar morphology, $\bfTheta=[0^\circ,23^\circ,37^\circ]$, loaded in the $\hat{\bfe}_3$ direction with the corresponding stress-strain curve. 

\section*{Movie S7.}
\textit{In situ} compression of the cubic morphology, $\bfTheta=[20^\circ,23^\circ,28^\circ]$, loaded in the $\hat{\bfe}_1$ direction with the corresponding stress-strain curve. 

\section*{Movie S8.}
\textit{In situ} compression of the cubic morphology, $\bfTheta=[20^\circ,23^\circ,28^\circ]$, loaded in the $\hat{\bfe}_2$ direction with the corresponding stress-strain curve. 

\section*{Movie S9.}
\textit{In situ} compression of the cubic morphology, $\bfTheta=[20^\circ,23^\circ,28^\circ]$, loaded in the $\hat{\bfe}_3$ direction with the corresponding stress-strain curve. 
\newpage
\bibliographystyle{plain}
\bibliography{Bib}